\documentclass[twocolumn]{aastex631}

\newcommand{\revII}[1]{\textrm{#1}}
\newcommand{\revIII}[1]{\textrm{#1}}
\newcommand{\revIV}[1]{\textrm{#1}}
\usepackage{bm}
\usepackage{csquotes}
\usepackage{subfiles}

\begin{document}

\title{Binary Analysis and Period Study of the Long-Period, High Mass Ratio Contact Binary KIC 7766185}

\author[0000-0002-4524-9497]{Conor M. Larsen}
\affiliation{Department of Physics and Astronomy,
University of Delaware, 
217 Sharp Lab, 
Newark, DE, 19716, USA}

\author[0000-0002-1913-0281]{Andrej Pr\v{s}a}
\affiliation{Department of Astrophysics and Planetary Science, 
Villanova University, 
800 E Lancaster Ave, 
Villanova, PA, 19085, USA}

\begin{abstract}

We present an in-depth photometric and spectroscopic study of the contact binary KIC 7766185. Spectroscopic observations were conducted \revII{on} the Mayall 4-m telescope at Kitt Peak National Observatory and used to extract radial velocities. Using the radial velocity measurements, \textit{Kepler} photometry, and \textit{Gaia} multi-color photometry, binary analysis was performed in \texttt{PHOEBE} to determine orbital and stellar parameters. The results classify KIC 7766185 as an A-type W UMa system and reveal one of the largest and most massive secondary stars in the sample of well studied W UMa systems. The two stars are in a shallow contact state, with a small fillout factor of $0.029\pm0.002$. Along with the binary analysis, we conduct a period study that reveals evidence for a cyclic variation in the eclipse timings.
\end{abstract}


\section{Introduction} \label{sec:intro}

Contact binaries are systems where two stars orbit so close that they share a common envelope. Of the contact binaries, the most common type is the W UMa type, \revIII{which consist of two, late-type (mostly of spectral type F, G and K), low mass stars which have come in contact and share a common, convective envelope \citep{dirk}. Early-type contact binaries, which consist of massive components which share a radiative envelope, do also exist \citep[e.g.][]{massive_contacts, Fabry_massive_contacts}.}

The origin and evolution of these objects are an active research area \citep[e.g.][]{recent_5, recent_4, recent_3, recent_2, recent_1}. The theoretical explanation for W UMa systems states that the stars come in thermal and geometric contact \citep{lucy}. It is believed that contact binaries start as detached binaries. Through angular momentum loss and subsequent orbital decay, the stars migrate towards each other until contact is achieved \citep{ruc}. Mechanisms for orbital contraction include tidal friction induced by Kozai-Lidov cycles through a distant third body and magnetic braking \citep[see][for an overview of possible formation mechanisms]{Eggleton_2012}.

Once the stars come in geometrical contact, they come in thermal contact with one another. Once in thermal contact, thermal energy is transferred between the two stars, forcing the temperature of the two objects to \revIII{nearly} equalize \citep{Lucy_models}.

W UMa stars have 3 subtypes based on temperature and size. A-type systems occur when the larger star is the hotter component, and W-type systems occur when the smaller star is the hotter component \citep{binn_subtype}. B-type W UMa stars are in geometric contact, but not thermal contact, which allows the two stars to have vastly different temperatures. In their study, \cite{Csizmadia} define B-type systems as those with a temperature difference greater than 1000 K. A 4\textsuperscript{th} subtype, proposed by \cite{Csizmadia}, divides the sample of W UMa stars by mass ratio. The H-type systems are high mass ratio systems with a mass ratio larger than 0.72.

Through eclipse timing variation (ETV) curves (also known as an O-C curve), orbital period changes can be inferred. The two major theoretical approaches to the evolution of contact binaries are angular momentum loss (AML) and the thermal relaxation oscillation (TRO) model. AML will cause the orbital period to decrease, driving the stars close and increasing the degree of contact \citep{ruc}. TRO predicts that contact systems oscillate between contact and semi-detached states. During the contact phase, mass transfers from the secondary to the primary leading to a period and orbital separation increase. The binary orbit increases until contact is broken. At this point, the system enters a semi-detached state with mass transferred from the primary to the secondary, leading to an orbital period and separation decrease \citep{LUCY_TRO, Flannery_TRO, Robertson_Eggleton_1977}. When the two theories are combined, AML can be strong enough to extend the contact phase and prevent the semi-detached phase of TRO from occurring \citep{ruc}. Since both theories predict period changes, period studies of contact binaries reveal the evolutionary status of the system. Additionally, cyclic variations in ETV curves are evidence of a potential third body or magnetic cycles in the system \revIV{\citep{Borkovits}. }

This paper \revII{discusses} the observations of the target star (Section \ref{sec:obs}), the radial velocity measurements (Section \ref{sec:RV meas}), any impact of extraneous light (Section \ref{sec:third light}), the binary analysis (Section \ref{sec:binary analysis}), the period study (Section \ref{sec:period study}), a comparison of the target to other individually studied W UMa stars (Section \ref{sec: context}) and finally a discussion of the results and a conclusion (Sections \ref{sec:discussion} and \ref{sec:conc}).

\section{Observations} \label{sec:obs}

This study focuses on the contact binary KIC 7766185. KIC 7766185 is a 12\textsuperscript{th} magnitude system of spectral type F3V \citep{frasca_data}. A review of previously determined parameters is presented in Table \ref{tab: lit values}. Note that the values for effective temperature and $\log g$ are estimates \revII{from} the average values of the two stars \citep[][]{KIC}. A comparison between the results of this study and previous works is discussed in Section \ref{sec:discussion}.

\begin{deluxetable}{|lcc|}
    \tablehead{\colhead{Parameter} & \colhead{Value} & \colhead{Reference}}
    \tablecaption{Previously Determined Parameters of KIC 7766185}
     \startdata 
      Distance [pc] & $934 \pm 9$ & 1 \\
        RA [$^{\circ}$] & $296.0164$ & 2 \\
        Dec [$^{\circ}$]& $43.4368$ & 2 \\
        Kepler Magnitude & 12.114 & 2 \\ 
        $T_{\text{eff}}$ [K] & 5977 & 2 \\
        $\log g$ [g cm$^{-3}$] & 4.280 & 2 \\
        Metallicity & -0.159 & 2 \\
        $E\left(B-V\right)$ & 0.077 & 2 \\
        Orbital Period [d] & 0.8354573(8) & 3 \\
        BJD$_{0}$ & $2454954.55\pm0.04$ & 3 \\
        Spectral Type & F3V & 4  \\ 
     \enddata
    \tablecomments{References: 1: \textit{Gaia DR3} 2078036917346313088, \cite{gaia}. 2: \textit{Kepler Input Catalog}, \cite{KIC}. 3: \textit{Kepler Eclipsing Binary Catalog}, \cite{Prsa_data}. 4: \cite{frasca_data}.}
    \label{tab: lit values}
\end{deluxetable}

\vspace{5 mm}

\subsection{Photometric Observations}\label{sec: photo}

\begin{figure}
    \centering
    \includegraphics[width=0.45\textwidth]{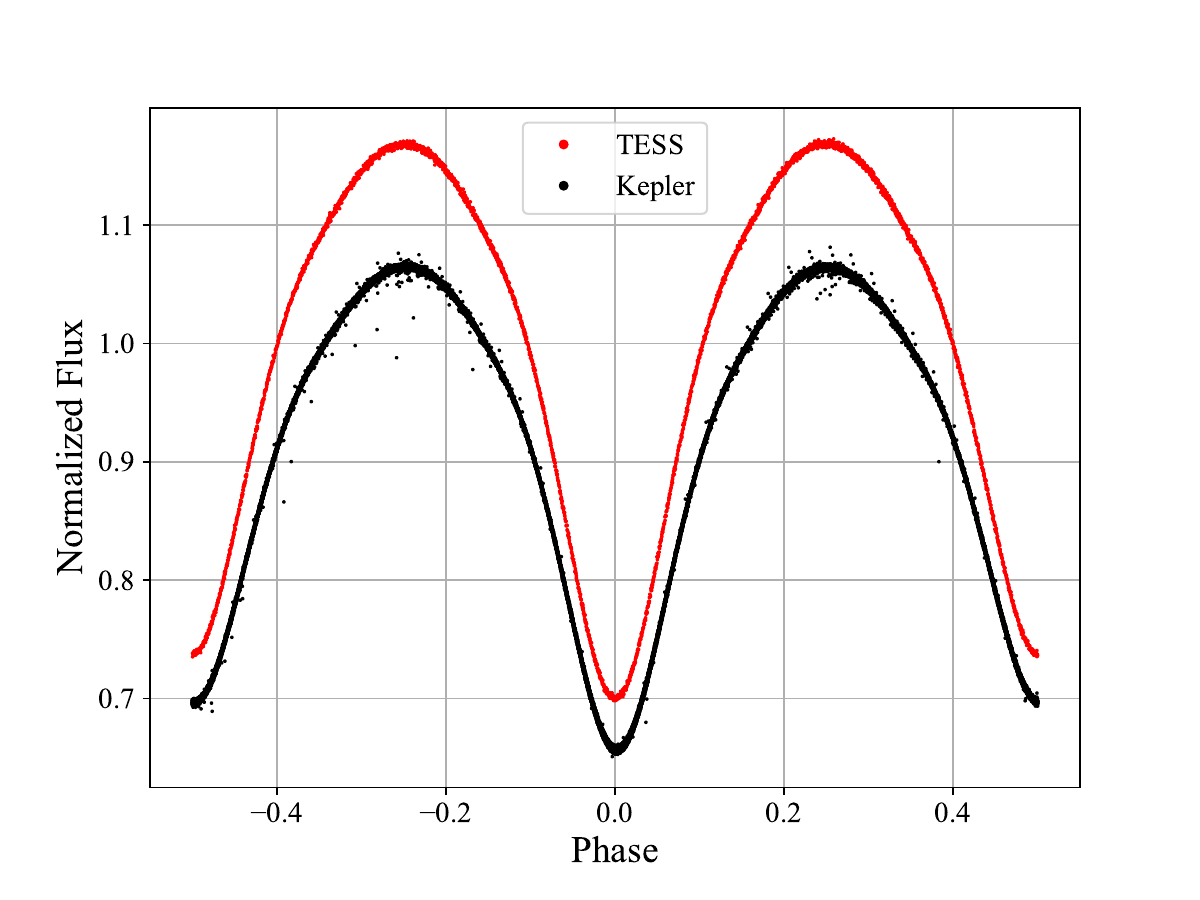}
    \caption{Phase-folded light curve of KIC 7766185 \revIII{from \textit{Kepler} (black) and \textit{TESS} Sector 55 (red). The light curves are normalized to their median values.}}
    \label{kepler lc}
\end{figure}

KIC 7766185 was observed by the \textit{Kepler Space Telescope} \citep{kepler} between 2009 and 2013 in Quarters 0 - 17. The observations were obtained at long cadence (30 minutes) in the \textit{Kepler} passband which covers a wavelength range of 400 - 850 nm. The \textit{Kepler} data were downloaded from the \textit{Kepler Eclipsing Binary Catalog} \citep[KEBC,][]{Prsa_data}. Detrending is conducted by the KEBC through iterative \revII{Legendre} polynomial fits to remove any intrinsic variability, such as stellar activity, and extrinsic variability, such as instrumental artifacts \cite[see][]{detrending}. The \textit{Kepler} phase-folded light curve is displayed in figure \ref{kepler lc}. The light curve has rounded tops and similar eclipse depths, both features of contact binaries. Outliers were removed using the outlier removal function in the period analysis software \texttt{Peranso} \citep{peranso}, which uses the LOWESS (LOcally-WEighted Scatterplot Smoothing) method. \revIII{The primary minimum for the \textit{Kepler} light curve was not centered exactly at a phase of zero. We adopted a new ephemeris of $T_{0} = 2454954.556508$ BJD to correct this slight shift.}

After removing outliers, the light curve contained $\sim$57,000 data points. Performing the binary analysis with this many measurements would be computationally expensive. The last piece of light curve processing conducted was to remove data points until the light curve consisted of 448 evenly spaced points. This greatly reduces the computational time while still maintaining dense coverage of the light curve. 

KIC 7766185 was also observed by the \textit{Transiting Exoplanet Survey Satellite} \citep[\textit{TESS},][]{TESS} during Sectors 14 and 15, from July 18 to September 10, 2019, and then in Sectors 54 and 55, from July 9 to September 1, 2022. Observations were conducted at 2 minute cadence at a wavelength range of 600 - 1000 nm yielding $\sim$50 days of observations. For this study, the binary modeling was conducted with the phase-folded \textit{Kepler} light curve. The \textit{TESS} observations were utilized for the period study (Section \ref{sec:period study}). \revIII{The phase-folded light curve for Sector 55 was downloaded from the \textit{TESS-EBs} catalog \citep[][]{TESS_EBs} and is displayed in figure \ref{kepler lc}.\footnote{The MAST doi for the \textit{TESS-EBs} catalog is: \dataset[10.17909/t9-9gm4-fx30]{https://dx.doi.org/10.17909/t9-9gm4-fx30}.}}

Multi-color photometric observations of KIC 7766185 were obtained by the \textit{Gaia Spacecraft} \citep{GAIA_spacecraft} between September 2014 - May 2017. The three \textit{Gaia} passbands are \textit{G} (green, 330 - 1050 nm), \textit{G\textsubscript{BP}} (blue, 330 - 680 nm) and \textit{G\textsubscript{RP}} (red, 630 - 1050 nm). \revII{The multi-color photometry was used to determine the temperatures of each star (Section \ref{sec: temp determination}).}

\subsection{Spectroscopic Observations}\label{sec: spectro}

\begin{figure}
    
    \includegraphics[width=0.45\textwidth]{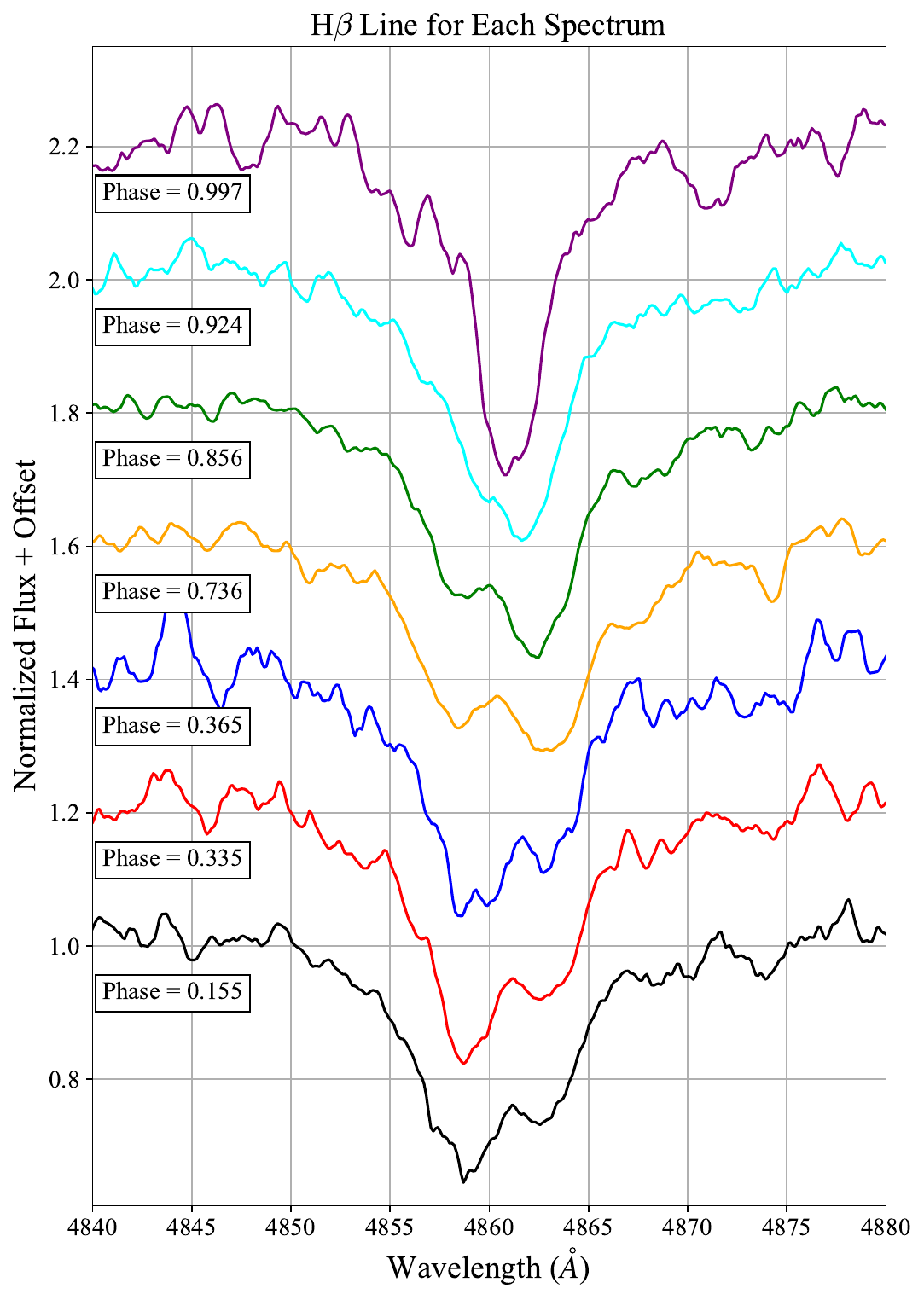}
    \caption{The H$\beta$ line for all seven spectra. The spectra are ordered by phase with an offset in normalized flux. The spectra were smoothed with the Savitzky–Golay filter to make both lines more apparent. We used a 3\textsuperscript{rd} order filter with a window length of 15.}
    \label{fig: example spectra}
\end{figure}

Spectroscopic observations of KIC 7766185 were obtained by the Kitt Peak Mayall 4-m telescope located at Kitt Peak National Observatory, Arizona. The observations were conducted with the 58-63 Echelle spectrograph (58.5 grooves/mm) and T2KB chip. The GG-420 filter was used, which covered a wavelength range of $\sim$4000 - 10000 {\AA} with a dispersion of $\Delta \lambda = 0.12$ \AA. A total of 7 spectra were obtained between September 26, 2010 and September 28, 2010, each with an exposure time of 600 s. The spectra were reduced and normalized in \texttt{IRAF} \citep{iraf} using the \texttt{echelle} package \citep{echelle}, and the \texttt{crreject} algorithm was used to remove cosmic rays from the data. Figure \ref{fig: example spectra} displays the H$\beta$ line for all 7 spectra. Two distinct peaks can be determined for all the spectra except the two obtained at a phase above 0.9, making this an SB2 system. Table \ref{Tab: spectra table} displays a list of the 7 spectra along with the corresponding Julian date and phase.

\begin{deluxetable*}{|c|c|c|c|}
    \tablehead{\colhead{Spectrum Number} & \colhead{BJD} & \colhead{Phase} & \colhead{Radial Velocity [km/s]}}
    \tablecaption{Spectroscopic Observations}
    
     \startdata 
      1 & 2455465.636 & 0.736 & $95 \pm 4$ \\
          &                  &       & $-204 \pm 5$ \\ \hline
        2 & 2455465.736 & 0.856 & $68 \pm 4$\\
          &                  &       & $-163\pm7$ \\ \hline
        3 & 2455465.793 & 0.924 & $-$\\
          &                  &       & $-$\\ \hline
        4 & 2455465.854 & 0.997  & $-33\pm4$\\
          &                 &        & $-33\pm4$ \\ \hline
        5 & 2455466.821 & 0.155 & $-156\pm5$\\
          &                  &       & $95\pm4$ \\ \hline
        6 & 2455467.807 & 0.335 & $-157\pm8$\\
          &                  &       & $112\pm4$ \\ \hline
        7 & 2455467.831 & 0.365 & $-140\pm9$ \\ 
          &                  &       & $102\pm4$ \\ 
     \enddata
    \tablecomments{List of spectra along with the barycentric Julian date and phase \revIII{of mid-exposure. Due to the 600 s exposure time, each value of mid-exposure time has an error of 0.003 and each value of phase has an error of 0.004.} Radial velocities are given in the fourth column. The top value is the primary component and the bottom value is the secondary component. Spectrum 3 was excluded due to severely blended broadening functions peaks.}
    \label{Tab: spectra table}
\end{deluxetable*}

\section{Radial Velocity Measurements} \label{sec:RV meas}

Radial velocity measurements were obtained using broadening functions due to the severe blending of \revII{spectral} lines in W UMa stars (see figure \ref{fig: example spectra}). Using a template \revII{spectrum with narrow lines}, the broadening function method \revIII{finds the convolution kernel that broadens the template to match the science spectrum.} The resulting broadening function can then be fit to extract the line shifts \cite[see][for a detailed description of the broadening function method]{BFs}. 

For a template, we downloaded a synthetic spectrum from the POLLUX database of synthetic stellar spectra \citep{pollux}. To employ broadening functions, we need a narrow line template which matches the atmospheric parameters of the stars. We downloaded an AMBRE spectrum \citep{AMBRE} with $T_{\text{eff}} = 6000$ K, $\log g = 4.00$ in cgs units, microturbulent velocity of 1 km/s and a metallicity of -0.25. These values generally match the literature estimates given in table \ref{tab: lit values}. While they may not match exactly (and the literature values themselves are only estimates) this is sufficient for the extraction of radial velocities (however, we did test additional templates, discussed below). 

\begin{figure}
    \centering
    \includegraphics[width=0.45\textwidth]{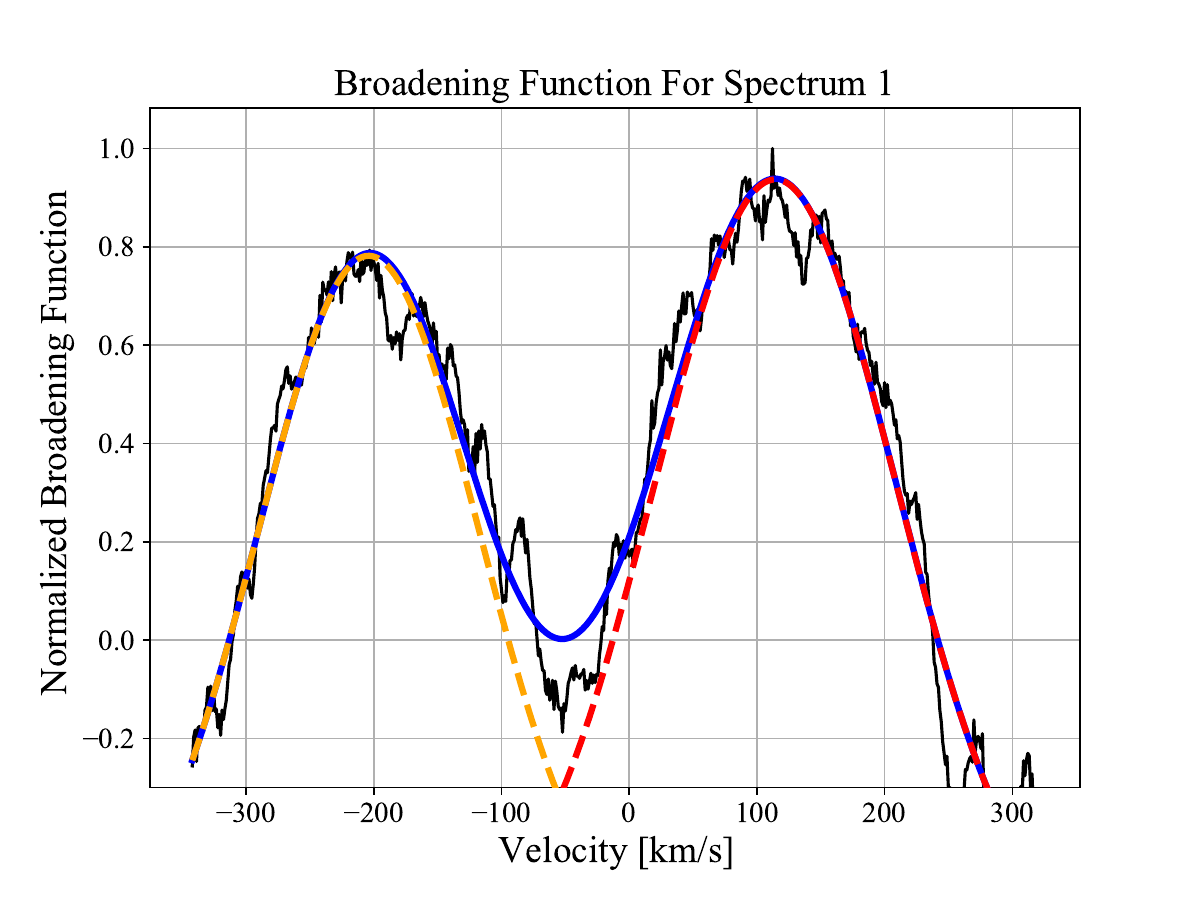}
    \caption{Normalized broadening function (black) computed with \texttt{SAPHIRES} along with Gaussian fits.}
    \label{bf example}
\end{figure}

We used the Python module \texttt{SAPHIRES} \citep[Stellar Analysis in Python for HIgh REsolution Spectroscopy,][]{saphires} to perform the broadening function analysis. \revIII{To avoid telluric interference, all spectra were trimmed to the range 4300 - 5400 {\AA}.} The broadening functions were then computed and fit to Gaussians. The broadening function along with the Gaussian fits for spectrum 1 is included in figure \ref{bf example} as an example. After applying a barycentric correction, the position of the center of the Gaussian represents the radial velocity of each star. The barycentric correction was computed with the \texttt{PyAstronomy} \citep{PyAstronomy}
function \texttt{helcorr}. 

\revIII{This process was conducted on the trimmed spectral range and also on the sub-ranges 4300 - 4900 {\AA} and 4900 - 5400 {\AA}. For the final radial velocity measurements, we took the average and standard deviation of the three measurements.} The radial velocities determined for each star for the seven spectra are listed in table \ref{Tab: spectra table}. Of special note are the results for spectra 3 and 4. As demonstrated in figure \ref{bf example}, ideally the broadening functions return two distinct peaks. Spectra 3 and 4 were both obtained at a phase greater than 0.9, and thus the broadening function peaks were blended together to form a single peak. For spectrum 3, obtained at a phase of 0.924, this means that extraction of radial velocities is impossible. Spectrum 4, however, was obtained at a phase of 0.997, during the primary eclipse. For spectrum 4, we fit the broadening function to a single Gaussian. This single value is then recorded as the radial velocity of both stars, thus being a measurement of the systemic velocity of the system.

For the error values on the radial velocity measurements, we took half of the velocity dispersion of the spectra. The velocity dispersion at a wavelength $\lambda$ is:

\begin{equation}
    \Delta v = \frac{c}{R} = c \left(\frac{\Delta \lambda}{\lambda}\right),
\end{equation}
where $c$ is the speed of light and $R$ is the resolving power of the spectrum. With a wavelength dispersion of $\Delta\lambda = 0.12$ \AA, the velocity dispersion at the center of the wavelength range used for the broadening function analysis ($\lambda_{\text{center}} = 4850$ \AA) is $\Delta v = 7.4$ km/s. Therefore we adopted a value of 4 km/s for the error \revIII{imposed by the spectral resolution. We have two error values, one from the standard deviation of the individual measurements and one from the spectral resolution. The errors reported in Table \ref{Tab: spectra table} are the larger of the two values.} 

In order to check the robustness of our fitting results, we also fit the broadening function peaks to Voigt profiles and quadratics and tried smoothing the broadening functions prior to fitting. All of these methods provided very similar results, within the adopted error values. We also re-measured the radial velocities with three additional template spectra to see if differences in the atmospheric parameters impacted the results. The differences were minor, always less than the adopted error values.

\section{Extraneous Light} \label{sec:third light}

Extraneous light is any additional flux added to the light curve from a source external to the binary system. It alters the amplitude of the light curve, thus impacting parameter determination. KIC 7766185 has two nearby sources with \textit{Gaia} \text{G} magnitudes of 17.8 and 16.7 (compared to 11.9 for KIC 7766185).\footnote{The \textit{Gaia} DR2 identifiers for the two nearby sources are 2078036913047484672 and 2078036917346312832.} Before conducting the binary analysis, we examined the \textit{Kepler} light curve further for any effects of extraneous light. 

\begin{figure}
    \gridline{\fig{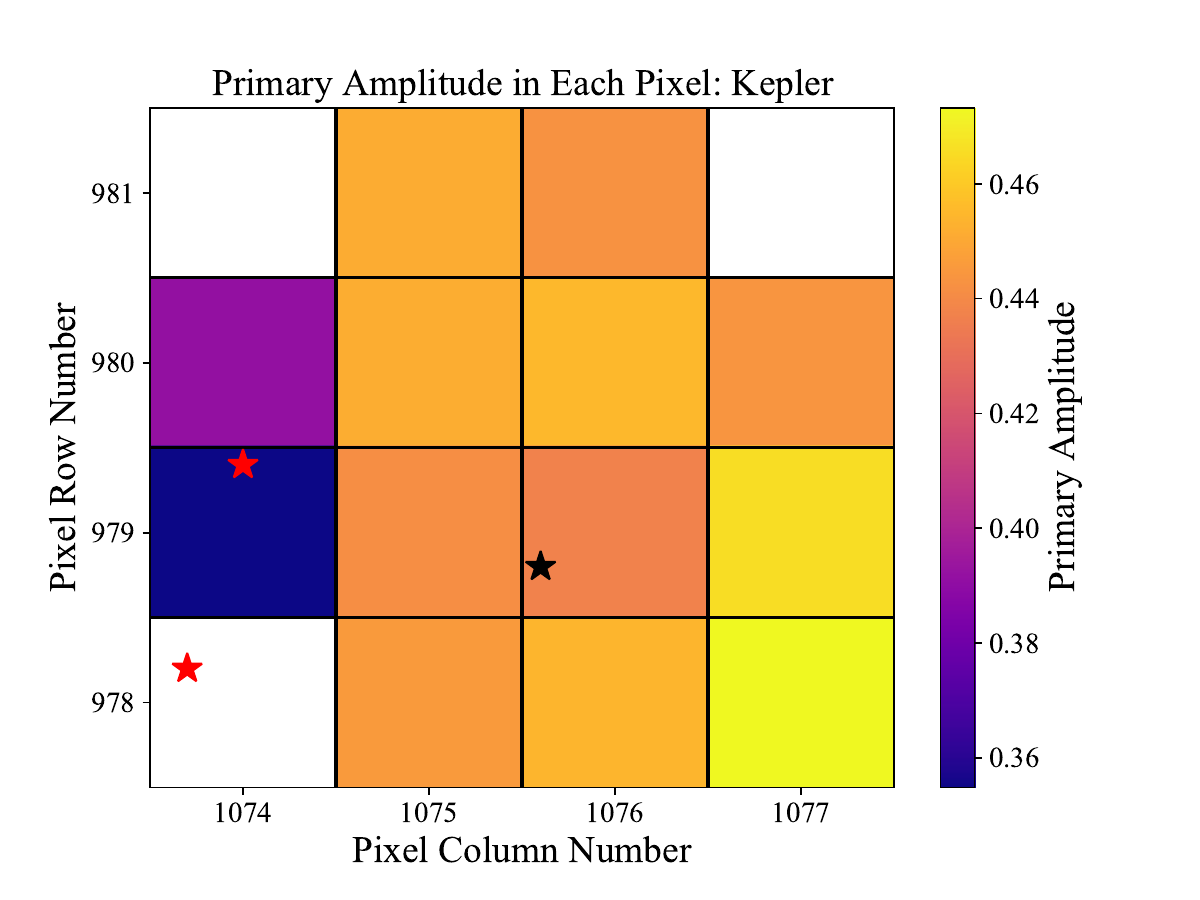}{\linewidth}{(a)}}
    \gridline{\fig{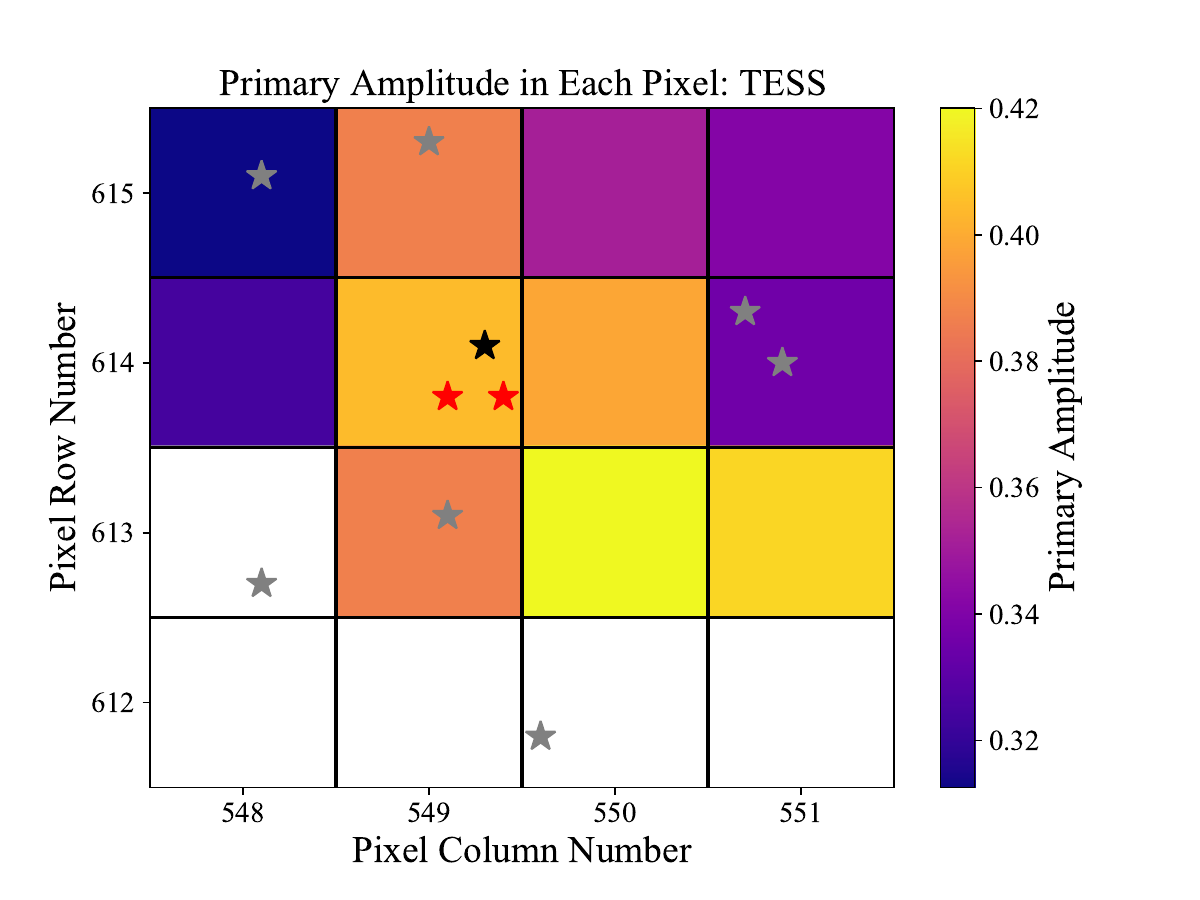}{\linewidth}{(b)}}
    \caption{(a) Amplitudes of the primary eclipse for each pixel in the \textit{Kepler} pipeline aperture for Quarter 8. The black star is the target KIC 7766185. The two red stars on the left are the faint \textit{Gaia} objects. The white pixels were not included in the \textit{Kepler} pipeline aperture for Quarter 8. (b) Amplitudes of the primary eclipse for each pixel in the \textit{TESS} pipeline aperture for Sector 15. The two red stars are the same objects in (a) and the gray stars are additional nearby objects. The white pixels were not included in the \textit{TESS} pipeline aperture for Sector 15.}
    \label{fig: l3 plot}
\end{figure}

To examine the effects of extraneous light, we used the Python package \texttt{Lightkurve} \citep{lightkurve} to extract a light curve for each pixel in the \textit{Kepler} pipeline aperture for Quarter 8. For each pixel in the \textit{Kepler} pipeline aperture, we fit the peak and primary minimum with Legendre polynomials and computed the primary amplitude. The primary amplitude in each pixel is displayed in figure \ref{fig: l3 plot}a. Two pixels (1074-979 and 1074-980) show a diminished amplitude. One of the external sources lies in the upper part of pixel 1074-979, explaining why these two pixels are most impacted. Due to this, we extracted a new light curve with a new aperture created by removing these two pixels. We found that removing these pixels had a very small impact on the light curve. The maximum difference between the light curve produced with the pipeline aperture and the new aperture is 0.00126 (in normalized flux). From this, we conclude that extraneous light has a negligible impact on the \textit{Kepler} light curve and can be safely ignored.

With a pixel size of ${\sim}21''$, \textit{TESS} is much more susceptible to extraneous light contamination. For Sector 15, the \text{TESS} pipeline aperture for KIC 7766185 contains seven additional stars. The two nearby \textit{Gaia} sources now lie in the same pixel as KIC 7766185. To examine the effect of extraneous light, we measured the primary amplitude in each pixel for \textit{TESS} Sector 15, using the same procedure discussed previously. The results are displayed in figure \ref{fig: l3 plot}b. The nearby sources lead to larger variations in the primary amplitude compared to \textit{Kepler}. \revIII{Figure \ref{kepler lc} also demonstrates that the amplitude of the \textit{TESS} light curve differs from the \textit{Kepler} light curve.} Due to this contamination, we only use the \textit{Kepler} light curve for binary modeling. The \textit{TESS} light curves were only used to measure times of minima, which is unaffected by extraneous light.

\section{Binary Analysis}\label{sec:binary analysis}

The binary analysis was conducted in \texttt{PHOEBE} \citep[PHysics Of Eclipsing BinariEs,][]{phoebe,PHOEBE_recent}. We used Markov Chain Monte Carlo (MCMC) sampling to determine the system parameters along with heuristic error estimates \citep[for a pedagogical review of MCMC sampling, see][]{MC_review}. Throughout the binary analysis, we adopted the orbital period given in the KEBC (see table \ref{tab: lit values}). The eccentricity was set to zero. Due to the close proximity of the stars in contact binaries, their orbits have circularized \citep{tidal_circ}. We find that the system is in a shallow contact state, connected by a very thin neck. In \texttt{PHOEBE}, the surface of the stars are discretized into triangles. In order to better resolve the thin neck connecting the stars, we increased the number of triangles to 6000 for all MCMC runs. All the detectors used in this study are photon-counting devices, therefore the intensity weighting was set to \texttt{photon} for the entire binary analysis \citep[see][]{photon_weight}. \revIV{Limb darkening is interpolated directly from \cite{Castelli_Kurucz} atmosphere tables \citep[see][]{photon_weight} and the gravity darkening coefficient of each component is 0.32. Irradiation/reflection effects are described by \cite{reflection} with a reflection fraction for each star of 0.6.}

Since the radial velocity curve only consists of six data points per star, when running the sampler they contribute little to the overall optimization if conducted with the light curve and radial velocity curve together. Therefore, we conducted two separate MCMC runs, one over the radial velocity curve, which we label the RV run, and one over the light curve, which we label the LC run. The parameters sampled in the RV run were the mass ratio ($q$), the projected semi-major axis ($a \sin i$), and the systemic velocity ($\gamma$). The RV run was ran with 16 walkers for \revIII{2405} iterations.\footnote{The number of iterations quoted throughout the paper is the total number of iterations minus the burn-in value. The burn-in value is the number of iterations needed to reach the equilibrium of the solver. These burn-in iterations are always discarded.\label{fn: iter}} Each walker explores the parameter space separately and therefore we obtain \revIII{2405} iterations for each walker. Once the RV run was complete (convergence of the sampler is discussed below), the LC run was conducted. For the LC run, the passband was set to \texttt{Kepler:mean}. The LC run sampled over the fillout factor ($f$), inclination ($i$), temperature ratio ($T_{2}/T_{1}$) and the mass ratio. The passband luminosity and primary temperature were also added as nuisance parameters. The errors reported on \textit{Kepler} photometric measurements are often underestimated, and therefore the noise nuisance parameter was added to the sampler to expand the errors on the individual measurements \citep[see][]{noise}. A set of prior distributions was also added to the LC run which included the posterior distributions determined in the RV run. A Gaussian prior centered at 6080 K with a standard deviation of 125 K was placed on the primary temperature to keep the walkers from using inappropriate temperatures. The final prior included a uniform distribution on the fillout factor ranging from 0 to 0.1 to keep the fillout factor from falling into the semi-detached state. With the priors and parameters set, the LC run was conducted with 48 walkers for 4578 iterations.

\begin{figure*}
    \gridline{\fig{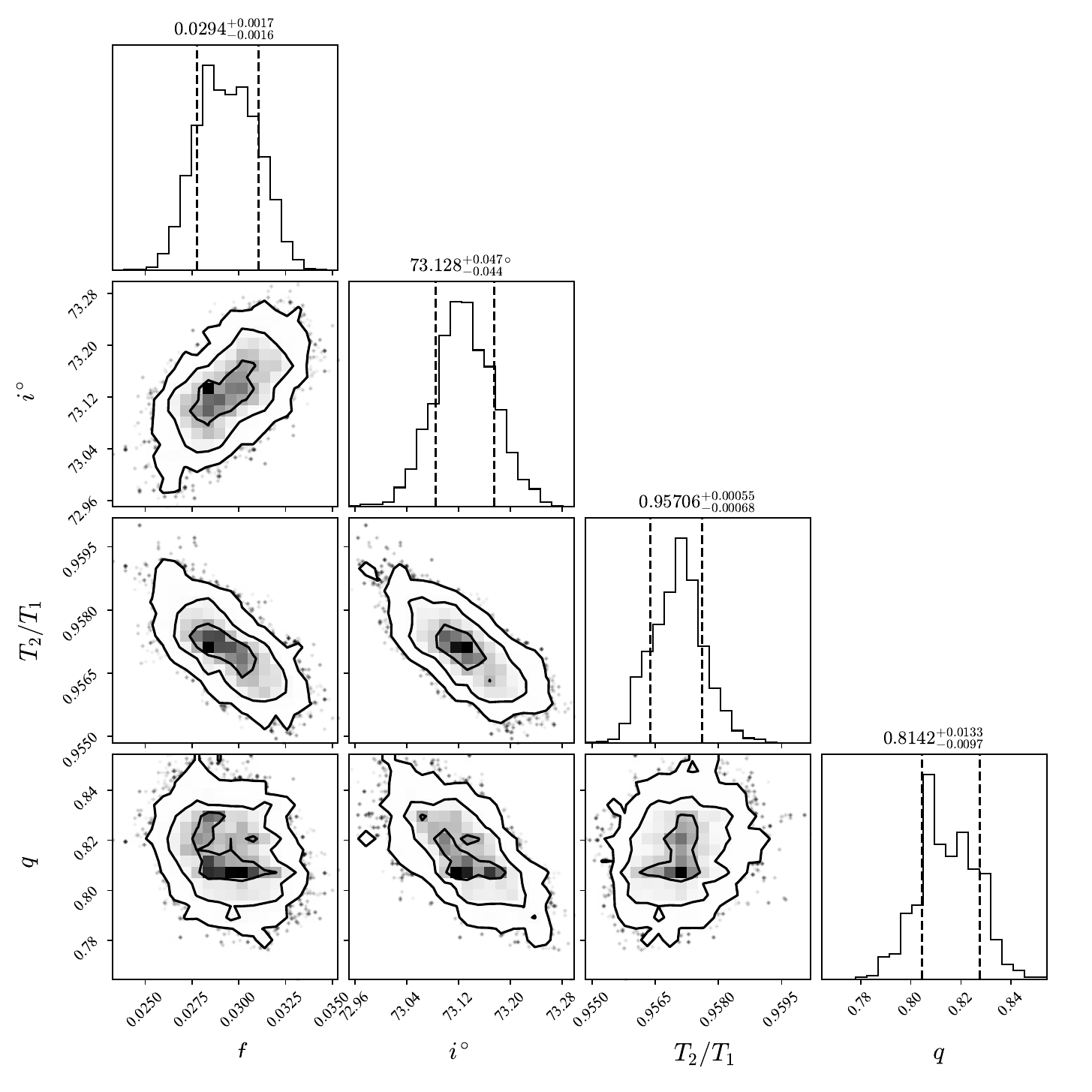}{0.5\textwidth}{(a)}
          \fig{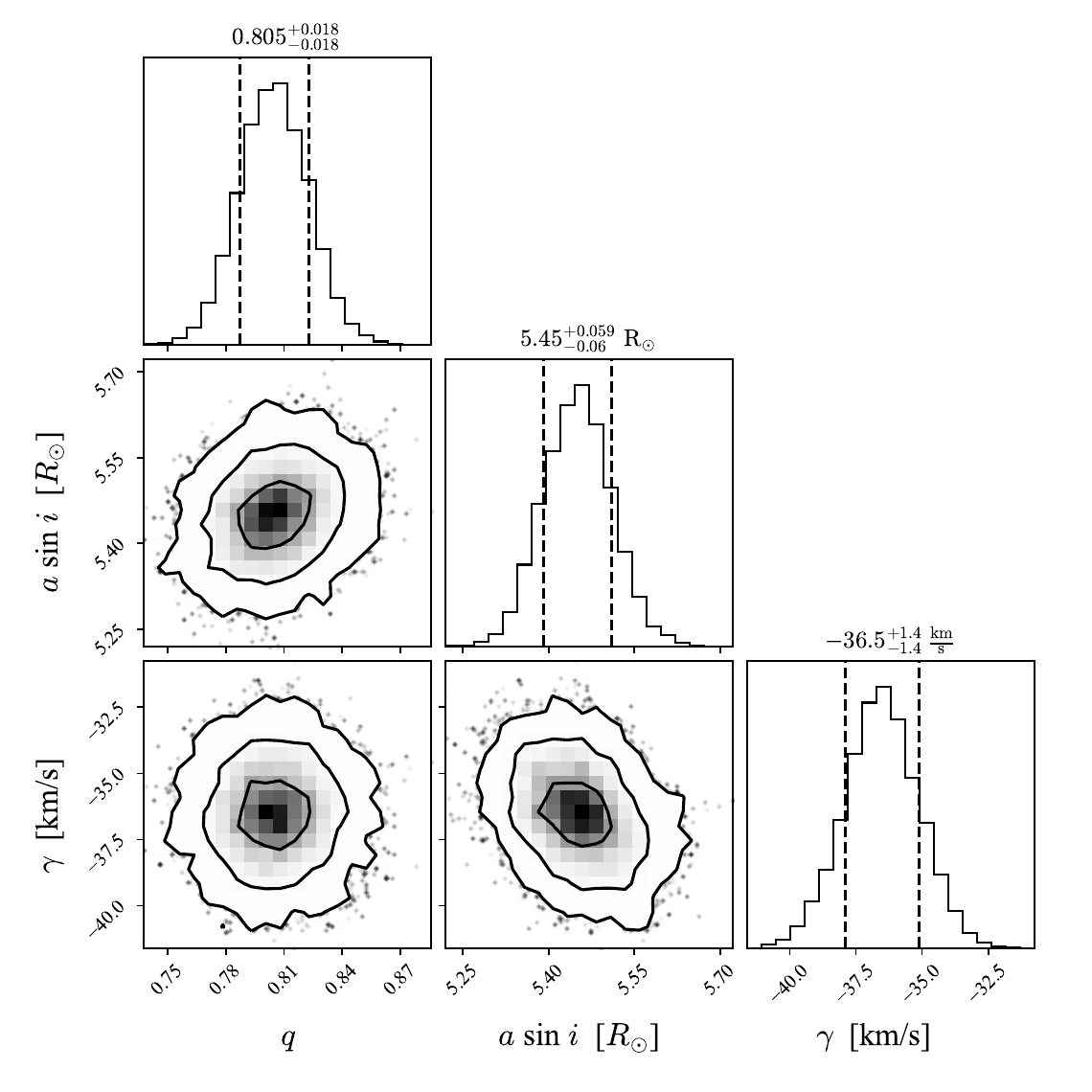}{0.4\textwidth}{(b)}}
          \gridline{\fig{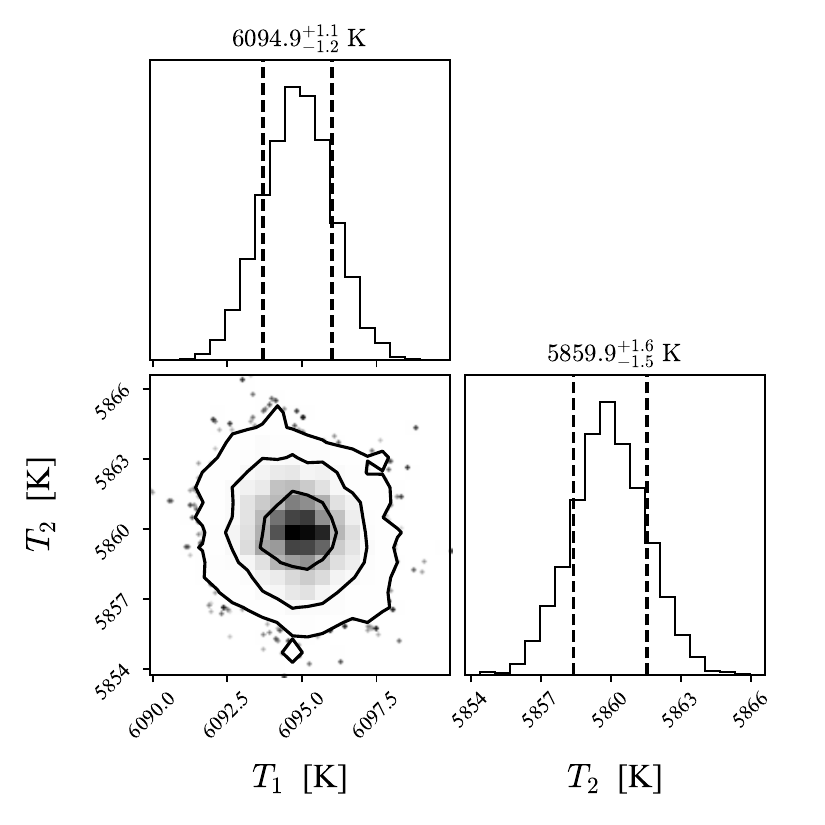}{0.4\textwidth}{(c)}}
    \caption{Distributions of the inferred parameters in the (a) LC run, (b) the RV run and (c) the temperature run. The vertical dashed lines in the one dimensional distributions represent the 68\textsuperscript{th} percentile. The three contours in the two dimensional distributions denote the 68\textsuperscript{th}, 95\textsuperscript{th} and 99.7\textsuperscript{th} percentiles.}
    \label{fig: dists}
\end{figure*}

Convergence occurs when the parameter space has been sampled thoroughly to produce appropriate parameter posteriors. A quantitative measure of convergence is the autocorrelation time, $\tau_{\text{ac}}$. The autocorrelation time is the number of iterations to produce an independent sample of the data. Typically, convergence of an MCMC run can be reached after a few autocorrelation times \citep{emcee}. Autocorrelation times are computed per parameter, giving a measure of how many iterations are required to independently sample for each parameter. Table \ref{Tab: autocorr times} lists the autocorrelation times along with $N_{\text{iters}} / \tau_{\text{ac}}$, where $N_{\text{iters}}$ is the total number of iterations, for all the sampled parameters for both MCMC runs. Both runs contained enough iterations to cover many autocorrelation times, and thus we can declare convergence. With the converged MCMC runs, the system parameters with error ranges can be calculated from the posterior distributions. The system parameters determined in this study are reported in table \ref{Tab: determined parameters}. Note that two mass ratios are determined, one from the light curve (the photometric mass ratio, $q_{\text{ph}}$) and one from the radial velocity curve (the spectroscopic mass ratio, $q_{\text{sp}}$). Both methods produce the same result of $0.81$. The corner plots displaying the distributions for the inferred parameters sampled in the LC run and RV run are presented in figure \ref{fig: dists}.

\begin{deluxetable}{|l c c|}
    \tablehead{\colhead{Parameter} & \colhead{$\tau_{\text{ac}}$ [iterations]} & \colhead{$N_{\text{iters}} / \tau_{\text{ac}}$}}
    \tablecaption{MCMC Autocorrelation Times}
    
     \startdata 
        RV run & & \\ \hline
        Mass Ratio & 35 & 69.0 \\
        $a\sin i$ & 35 & 68.1 \\ 
        Systemic Velocity & 48 & 50.4 \\ \hline
        LC run & & \\ \hline
        Fillout Factor & 253 & 18.1 \\
        Inclination & 268 & 17.1 \\ 
        Temperature Ratio & 295 & 15.5 \\ 
        Passband Luminosity & 218 & 21.0 \\
        Primary Temperature & 318 & 14.4 \\
        Temperature Ratio & 295 & 15.5 \\
        Mass Ratio & 216 & 21.2 \\
        Noise Nuisance Parameter & 213 & 21.5 \\ \hline
        Temperature run & & \\ \hline
        Primary Temperature & 23 & 54.7 \\
        Secondary Temperature & 36 & 35.2 \\ 
        Passband Luminosity & 39 & 32.9 \\ \hline
     \enddata
    \tablecomments{The autocorrelation times and the number of iterations over the autocorrelation time for each parameter sampled in the MCMC runs. As mentioned in footnote \ref{fn: iter}, the number of iterations used in the calculation of column three is the number of iterations after the burn-in period.}
    \label{Tab: autocorr times}
\end{deluxetable}

\begin{deluxetable}{|l c|}
    \tablehead{\colhead{Parameter} & \colhead{Value}}
    \tablecaption{System Parameters Determined for KIC 7766185}
    
     \startdata 
        $f$ & $0.029\pm0.002$ \\
        $i$ [$^{\circ}$]  & $73.13^{+0.05}_{-0.04}$ \\
        $T_{2}/T_{1}$ & $0.9571^{+0.0006}_{-0.0007}$ \\ 
        $q_{\text{ph}}$ & $0.81\pm0.01$ \\
        $R_{\mathrm{equiv,\:1}}$ [$R_{\odot}$]& $2.29^{+0.02}_{-0.03}$\\
        $R_{\mathrm{equiv,\:2}}$ [$R_{\odot}$]& $2.07\pm0.03$ \\ 
        $q_{\text{sp}}$ & $0.81\pm0.02$ \\
        $a\sin i$ [$R_{\odot}$]& 5.45 $\pm$ 0.06  \\
        $\gamma$ [km/s] & $-37\pm1$ \\
        $a$ [$R_{\odot}$] & $5.70\pm0.06$ \\
        $M_{1}$ [$M_{\odot}$]& $1.97\pm0.06$\\
        $M_{2}$ [$M_{\odot}$]& $1.58\pm0.06$ \\
        $T_{1}$ [K]& $6095\pm125$\\
        $T_{2}$ [K]& $5860\pm125$ \\
     \enddata
    \tablecomments{System parameters of KIC 7766185 determined from this study. The error ranges represent the 68\textsuperscript{th} percentile. The only exceptions are the effective temperatures. The errors for these two quantities are discussed in Section \ref{sec: temp determination}.}
    \label{Tab: determined parameters}
\end{deluxetable}

\begin{figure*}
    \gridline{\fig{model_LC_Curve.pdf}{0.4\textwidth}{(a)}
          \fig{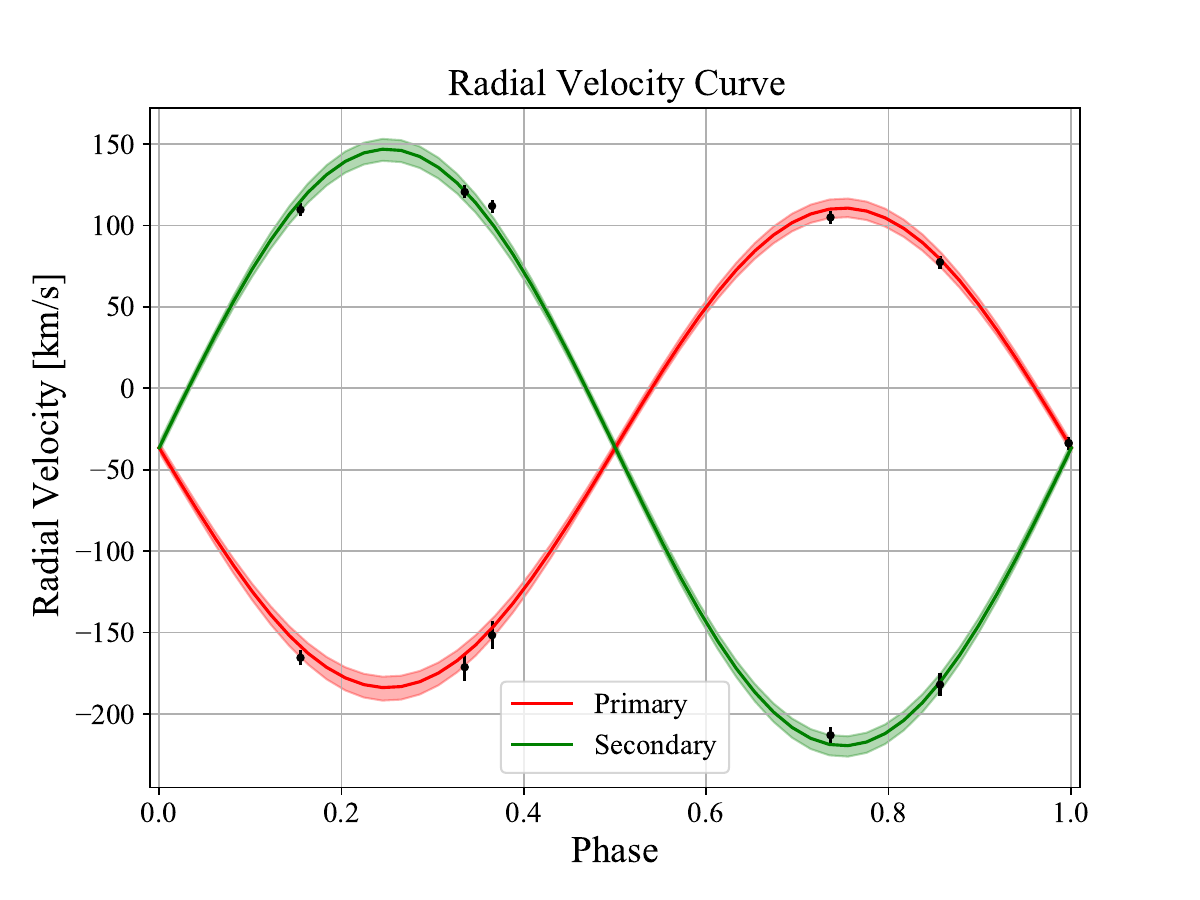}{0.4\textwidth}{(b)}}
          \gridline{\fig{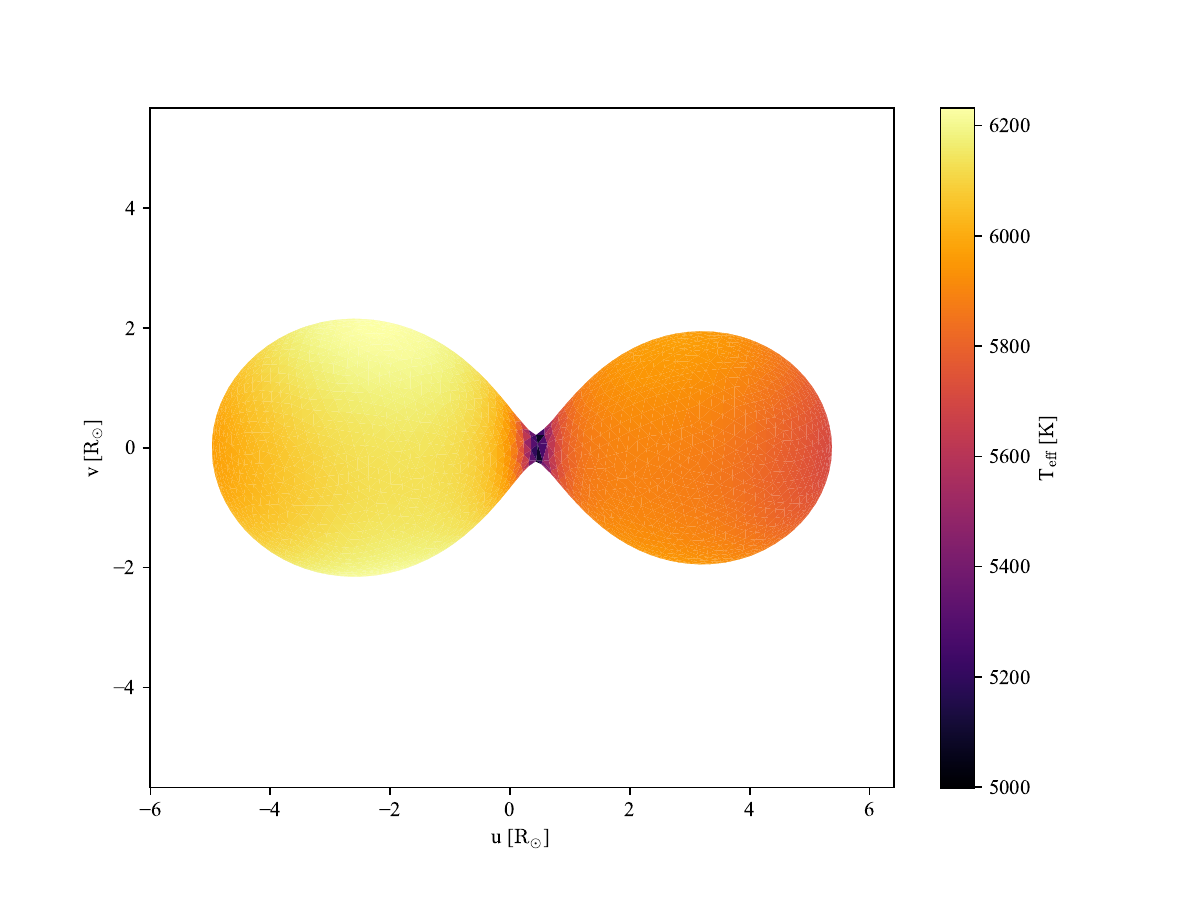}{0.4\textwidth}{(c)}
          \fig{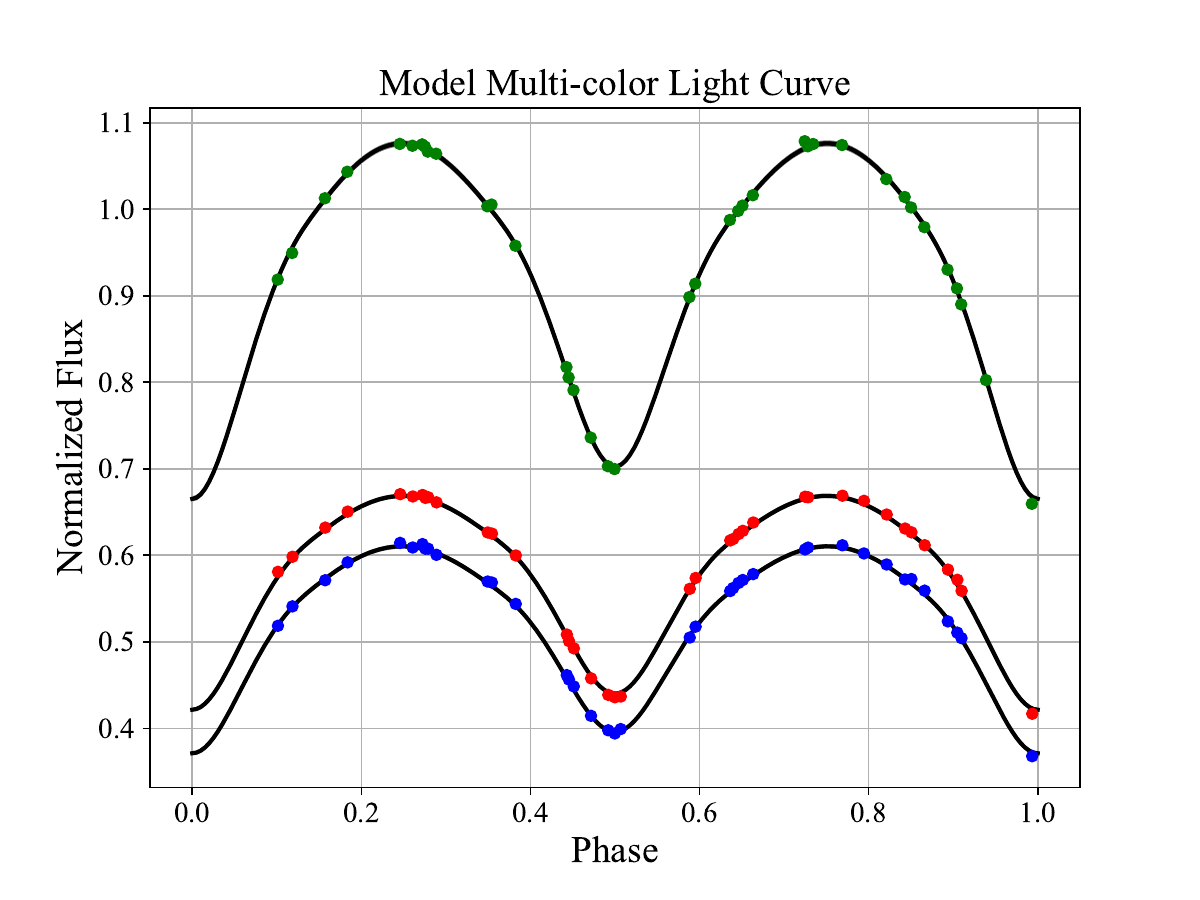}{0.4\textwidth}{(d)}}
    \caption{(a) Model light curve plotted on the \textit{Kepler} observations. (b) Model radial velocity curve along with the radial velocity measurements. (c) Mesh plot graphically displaying the system face on. (d) Model multi-color light curve along with the \textit{Gaia} observations. The three light curves in (d) are normalized to the median of the \textit{G}-mag measurements.}
    \label{fig: models}
\end{figure*}

With these parameters determined, a model light curve, radial velocity curve and mesh plot were created. For the model light curve and radial velocity curve we drew 100 samples from the posterior distributions of the MCMC runs. With these 100 sets of system parameters we ran 100 forward models and calculated the median and 3$\sigma$ spread. The resulting model light curve, radial velocity curve and mesh plot are displayed in figure \ref{fig: models}. The mesh plot is computed with the mean values reported in table \ref{Tab: determined parameters} (including the effective temperatures, discussed in the following section). The mesh plot, along with the small value for the fillout factor, reveals that KIC 7766185 is in a shallow contact state, connected via a thin neck with a radius of about 0.25 $R_{\odot}$.

\subsection{Temperature Determination}\label{sec: temp determination}

With the \textit{Gaia} multi-color photometry, the effective temperatures of the two stars can be determined. The three \textit{Gaia} light curves were normalized to the median of the \textit{G}-mag measurements. We then adopted the values determined in the LC run and conducted a third MCMC run. The run sampled for the primary and secondary temperatures. The passband luminosity was again added as a nuisance parameter. Uniform priors were set on the temperatures extending from 4000 - 7000 K. We had to update the \textit{Gaia} passband transmission functions (PTFs) in \texttt{PHOEBE}. The \textit{Gaia} PTFs in \texttt{PHOEBE} are the original pre-launch PTFs. There are significant differences between the pre-launch and \textit{Gaia} EDR3 PTFs \citep{EDR3_PTF}. When attempting the temperature determination with the pre-launch PTFs, we could not obtain satisfactory fits to the measurements. This was corrected after updating to the \textit{Gaia} EDR3 PTFs, which were downloaded from the \textit{Gaia} website.\footnote{\url{https://www.cosmos.esa.int/web/gaia/edr3-passbands}.} The MCMC run was conducted with 16 walkers for 1273 iterations. As done previously, we checked the autocorrelation times to test for convergence. The autocorrelation times and $N_{\text{iters}} / \tau_{ac}$ for the three sampled parameters in the temperature run are displayed in table \ref{Tab: autocorr times}. For all three parameters, we covered many times the autocorrelation time. The posterior distributions for the primary and secondary temperatures are displayed in figure \ref{fig: dists}. Model light curves were created in the three filters using the same procedure discussed in the previous section. The resulting model light curves are displayed in figure \ref{fig: models}. The efffective temperatures are $T_{1} = 6095$ K and $T_{2} = 5860$ K. The error ranges determined from MCMC are small, between 1-2 K. However, the atmosphere models used in \texttt{PHOEBE} are interpolated over a grid with a temperature spacing of 250 K. We find it more appropriate to use half this grid spacing as the error range for the effective temperature, thus in table \ref{Tab: determined parameters}, the error values are listed as 125 K. With these temperatures determined from the \textit{Gaia} multicolor photometry, the temperature ratio is $T_{2}/T_{1} = 0.96$, in close agreement with the value determined from the \textit{Kepler} light curve.

\section{Period Study} \label{sec:period study}

The \textit{Kepler} and \textit{TESS} observations together provide a baseline of ${\sim}15$ years to search for orbital period variations. Period variations are most easily inferred through eclipse timing variation (ETV) curves. The ETV value for a particular eclipse is the difference between the observed and predicted time of eclipse:

\begin{equation}\label{eq: ETV}
    \text{ETV} = T_{\text{obs}} - \left(T_{0} + PE\right),
\end{equation}
where $T_{\text{obs}}$ is the measured time of mid-eclipse, $T_{0}$ is the ephemeris, $P$ is the orbital period and $E$ is the epoch of the particular eclipse. For the \textit{TESS} observations, we downloaded the individual sectors using the Python package \texttt{Lightkurve}. We then imported these light curves into the period analysis software \texttt{Peranso} and measured the time of minimum for each primary eclipse using polynomial fitting. 

\begin{figure}
    \centering
    \includegraphics[width=0.45\textwidth]{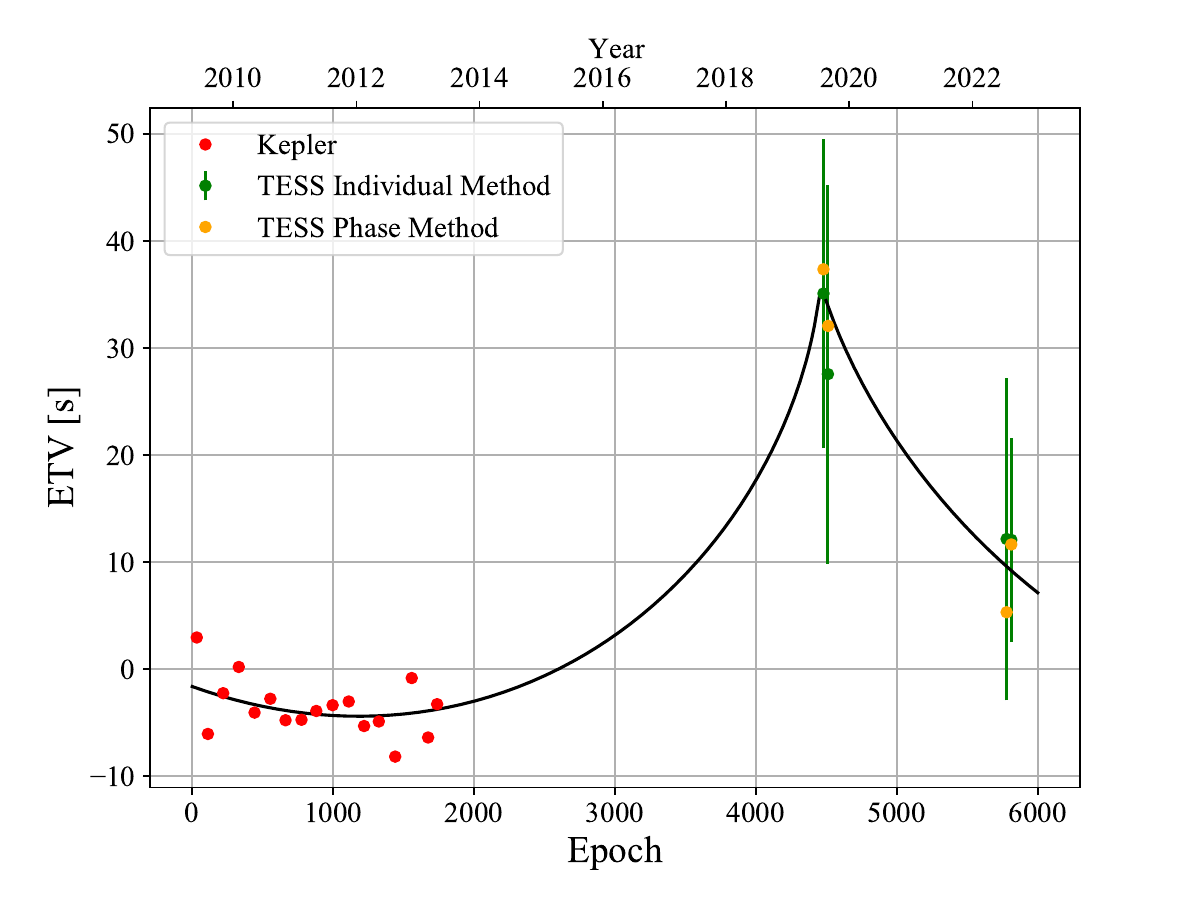}
    \caption{ETV curve for Kepler 7766185. The red and orange points are the \textit{Kepler} and \textit{TESS} ETVs determined through the phase method. The green points are the averages and standard deviations of the individual measurements over each \textit{TESS} sector. The black curve is a third body fit with an eccentricity of $e_{3}\sim0.95$ and a period of $P_{3}\sim6000$ days.}
    \label{fig: ETV curve}
\end{figure}

The 30 minute cadence of \textit{Kepler} makes precise measurements of times of minima for individual eclipses difficult. Instead, using \texttt{Lightkurve}, we downloaded each individual \textit{Kepler} quarter and phased the entire quarter. With the phased light curve, we smoothed the measurements using the LOWESS method and calculated the phase of minimum via polynomial fitting. This method allows us to combine many eclipses to improve the precision of the computed ETV value. We also performed this same method on the four \textit{TESS} sectors to check that the results are consistent with the individual eclipse measurements. Figure \ref{fig: ETV curve} displays the ETV curve. For the individual \textit{TESS} measurements, we took the average and standard deviation over each \textit{TESS} sector. The averaged individual measurements are in agreement with the results from the phase method. 

Cyclic variations in ETV curves can indicate the presence of a third body. It appears that a cyclic variation is present in the ETV curve, but without a full period we cannot say with certainty. Using the equations in \cite{ETV_third_body} plus an offset term, we fit the the ETV curve with a third body solution with a large eccentricity of $e_{3}\sim0.95$ and a period of $P_{3}\sim6000$ days. At this point, we do not wish to declare the existence of a third body, the fitting is merely to demonstrate that a third body solution is possible. As future \textit{TESS} sectors are released, more observations will allow for better constraints on the nature of the ETV curve.

\section{KIC 7766185 in Context}\label{sec: context}

\begin{figure}
    \centering
    \includegraphics[width=0.45\textwidth]{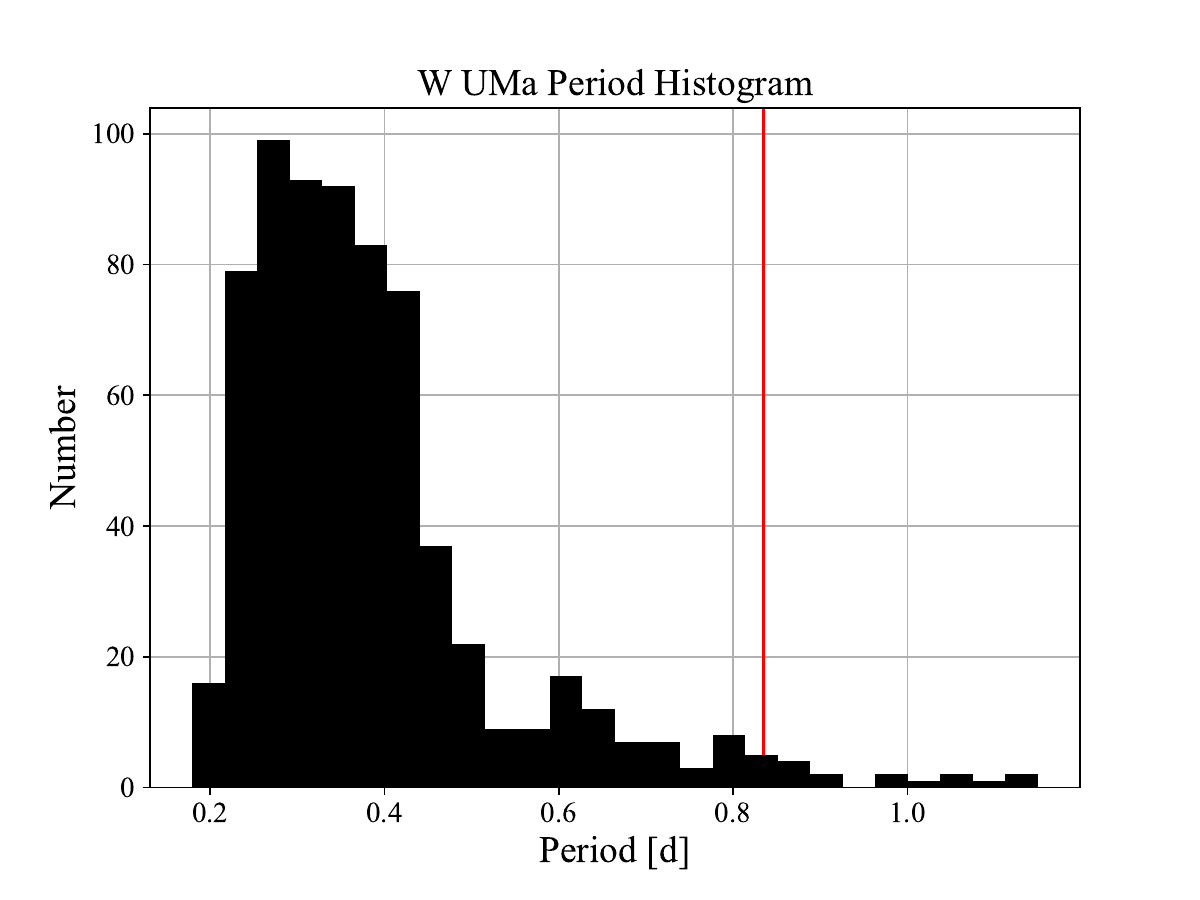}
    \caption{Period histogram for the systems in the \textit{WUMaCat} database. The red line denotes the period of KIC 7766185.}
    \label{fig: period hist}
\end{figure}

In this section we compare KIC 7766185 to the collection of other individually studied W UMa stars. The system parameters for the other systems were downloaded from the \textit{WUMaCat} database \citep{700_WUMas}, which contains data for around 700 individually studied W UMa stars. Of first note is that KIC 7766185 is a long period system. Figure \ref{fig: period hist} displays the period histogram for W UMa stars. The period of KIC 7766185 (0.835 days) is in the 97.7\textsuperscript{th} percentile of W UMa orbital periods. While figure \ref{fig: period hist} only displays the data for the ${\sim}700$ sources in \textit{WUMaCat}, the distributions formed from the much more numerous samples of \textit{LAMOST} and \textit{CRTS} also show similar distributions \citep[see][]{700_WUMas}.

\begin{figure*}
    \gridline{\fig{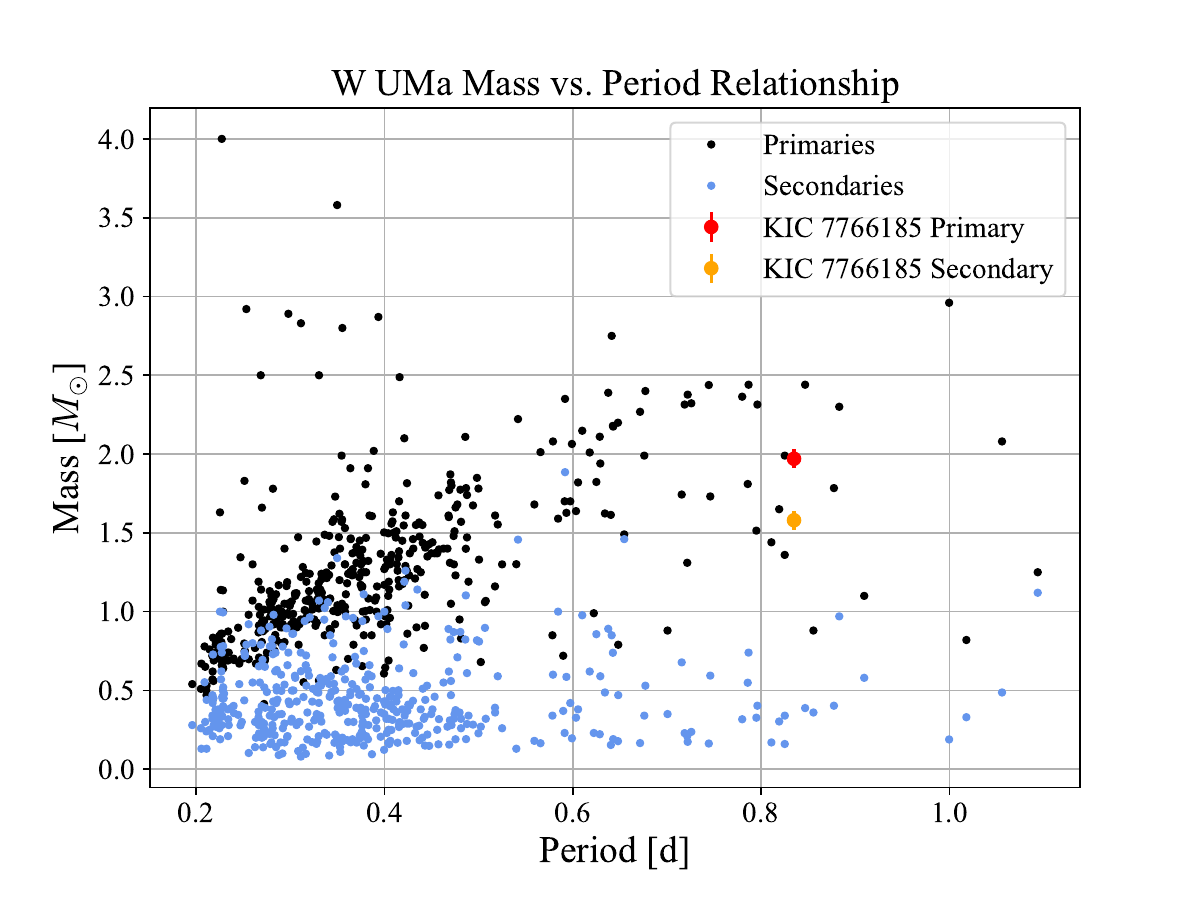}{0.4\textwidth}{(a)}
          \fig{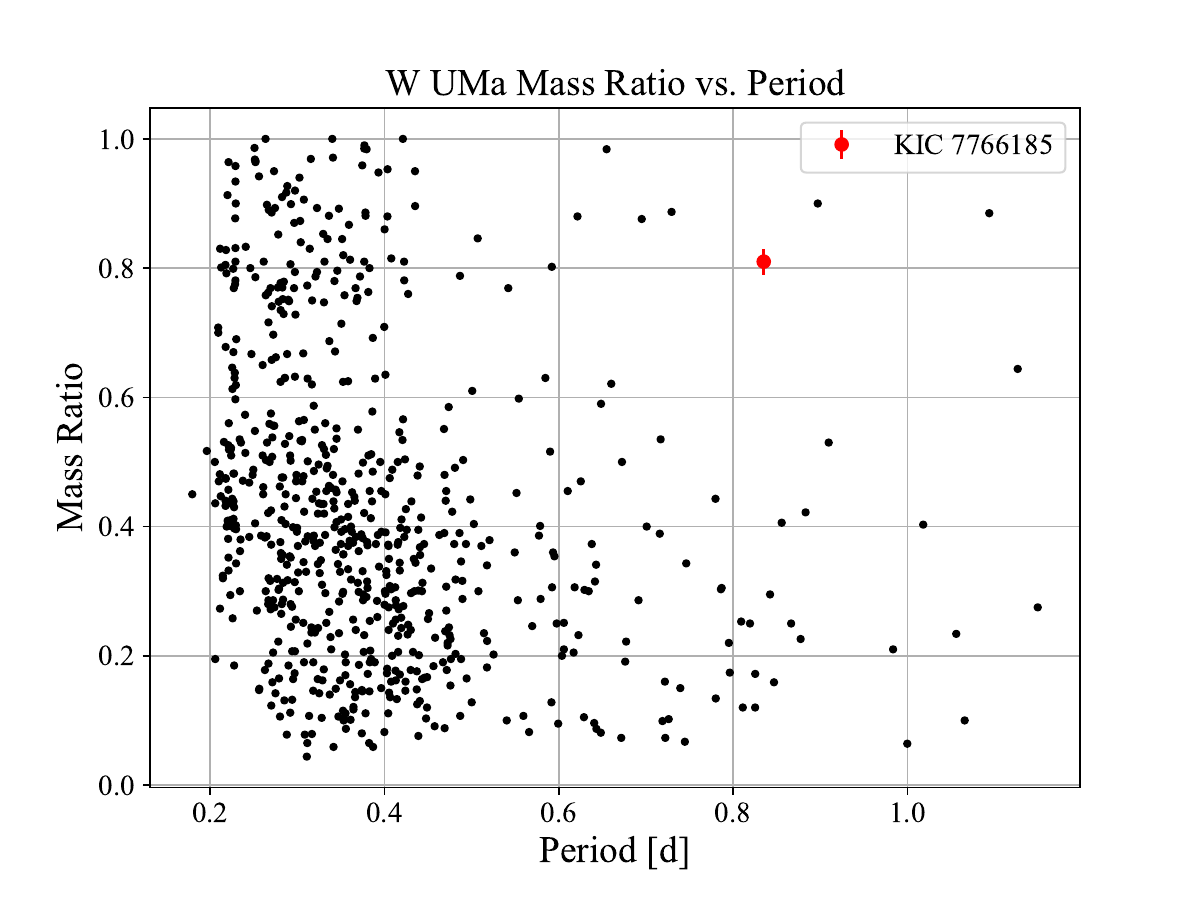}{0.4\textwidth}{(b)}}
    \gridline{\fig{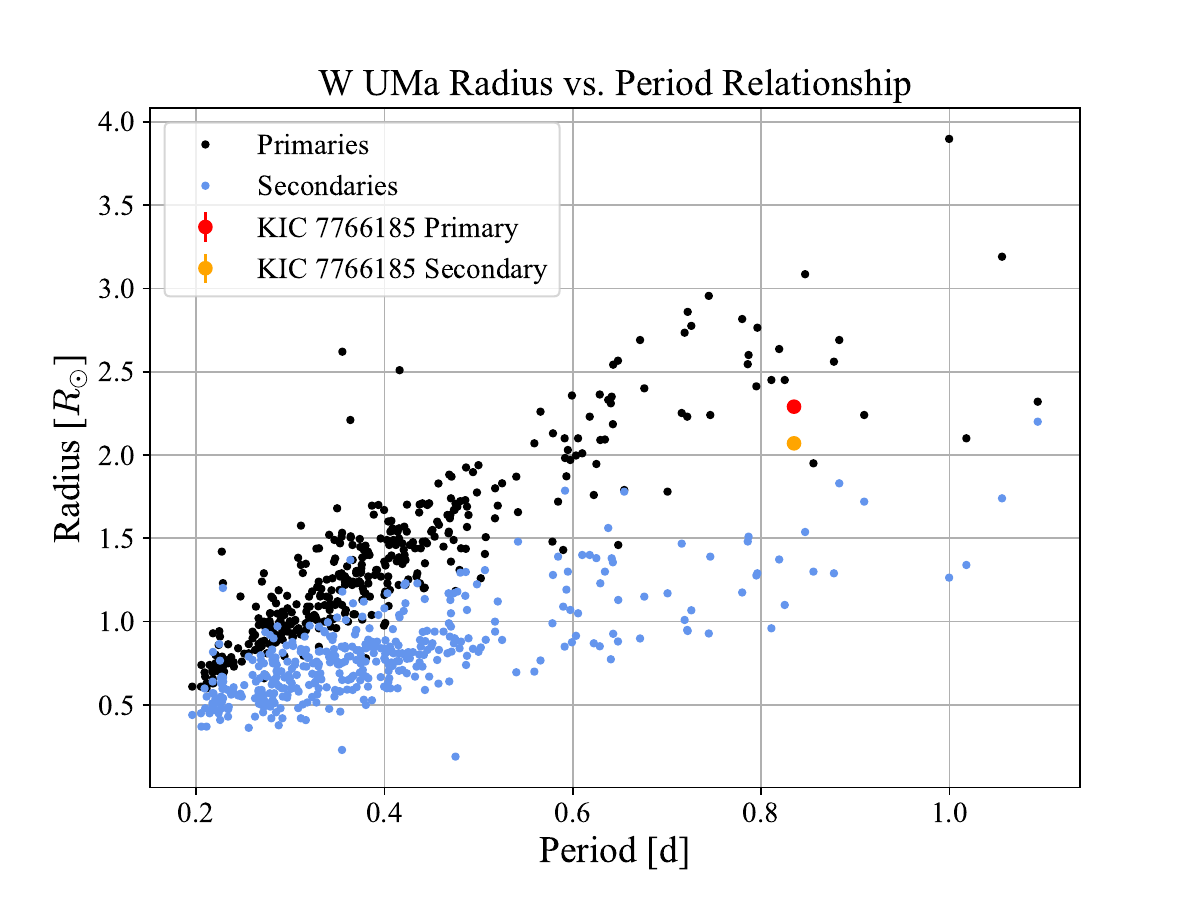}{0.4\textwidth}{(c)}
          \fig{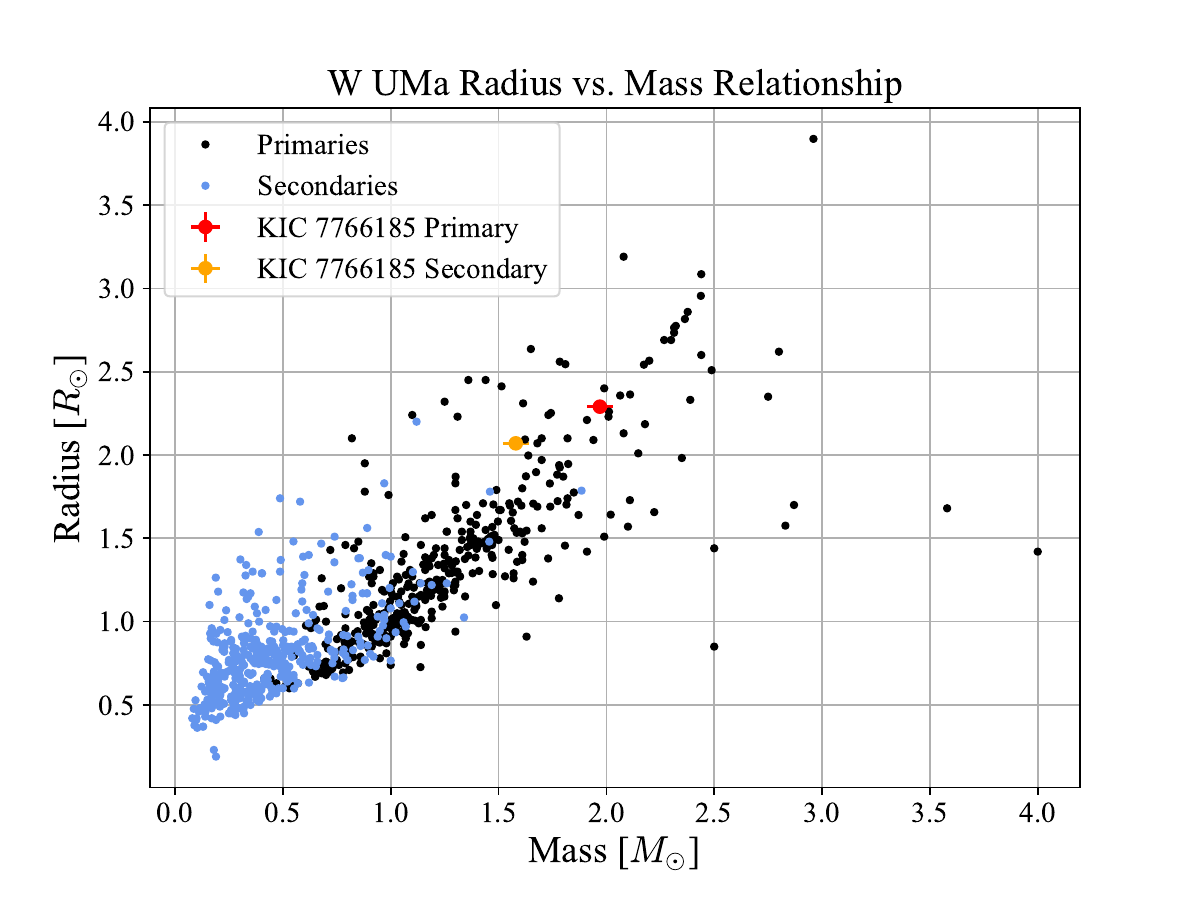}{0.4\textwidth}{(d)}}
    \caption{(a) The mass vs. period relationship for both the primary and secondary stars. (b) The mass ratio vs. period relationship. (c) The radius vs. period relationship for both the primary and secondary stars. (d) The radius vs. mass relationship for both the primary and secondary stars.}
    \label{fig: Stats}
\end{figure*}

Next we compare the masses and radii of each component of KIC 7766185 to the rest of the W UMa sample.  Figure \ref{fig: Stats}a plots the mass vs. period for the primaries and secondaries in the \textit{WUMaCat} sample. From this plot we find that the primary of KIC 7766185 is less massive and the secondary is more massive when compared to other systems of a similar orbital period. Figure \ref{fig: Stats}b displays the mass ratio vs. period relationship. Low period systems, below $\sim$0.5 days, cover the entire range of mass ratios while being biased towards low values. Long period systems, on the other hand, mostly have mass ratios below 0.6. The mass ratio of KIC 7766185 is 0.81 and, as indicated by figure \ref{fig: Stats}b, has one of the largest mass ratios when compared to the long period systems. Compared to the entire sample, KIC 7766185 is in the 90.1\textsuperscript{th} percentile of W UMa mass ratios. From these two mass plots, we learn that the distribution of the total mass equally into the two components is unlike other long period systems. In figure \ref{fig: Stats}c we plot the radius vs. period relationship. Just as with the masses, the two components of KIC 7766185 have near equal radii, which again is uncommon for long period systems. \revIII{Since there is a tight correlation between the mass ratio and radius ratio for contact systems, the high radius ratio is expected from the high mass ratio.} We also find that the secondary is enlarged and the primary undervalued in radius when compared to other W UMa systems of similar orbital period. Figure \ref{fig: Stats}d plots the radius vs. mass relationships. \revIII{Both stars are consistent with the mass-radius relation for other well studied W UMa systems}. KIC 7766185 is a long period system with similar components. It is an outlier when compared to other long period systems having a smaller, less massive primary and a larger, more massive secondary. From these plots we find that KIC 7766185 contains one of the largest and most massive secondaries, lying in the 99.8\textsuperscript{th} percentile of secondary radii and the 98.9\textsuperscript{th} of secondary masses.

\begin{figure*}
    \gridline{\fig{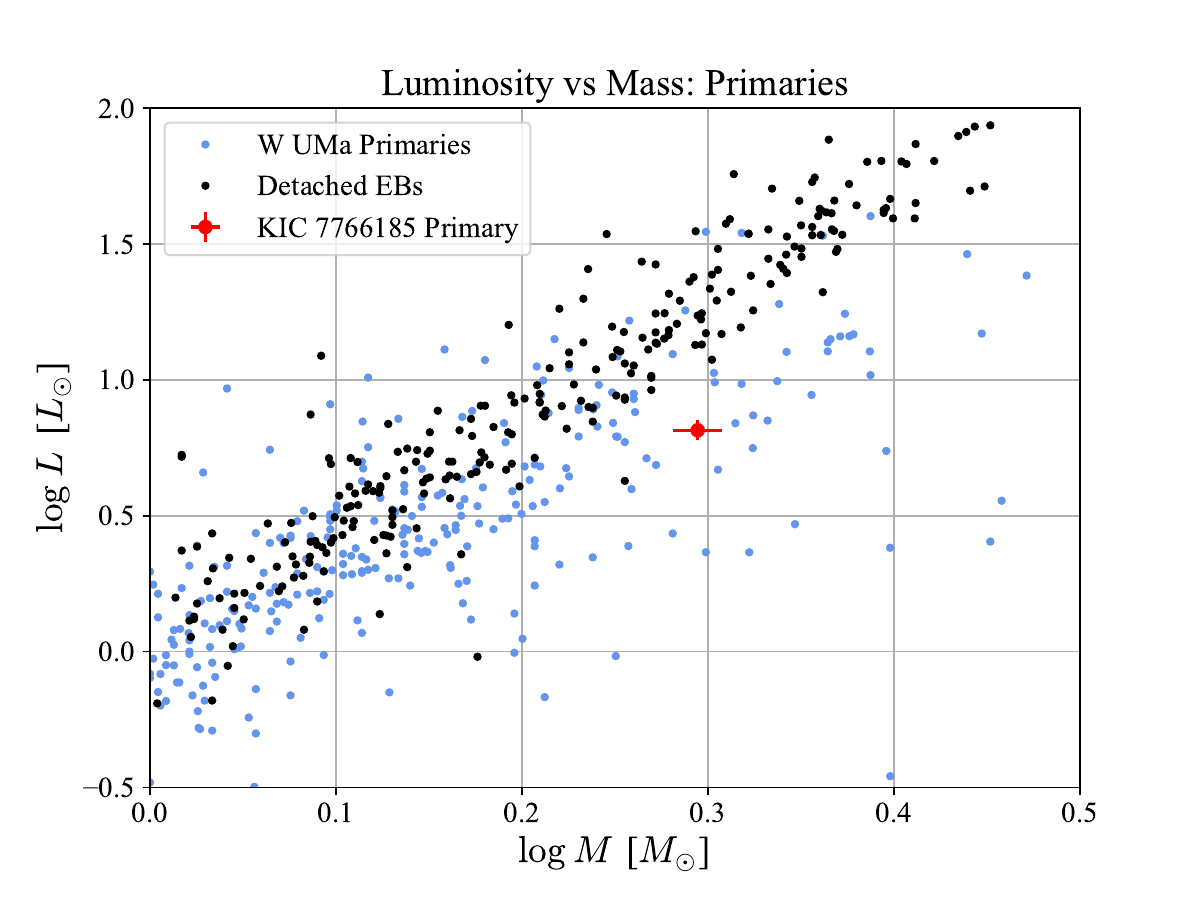}{0.4\textwidth}{(a)}
          \fig{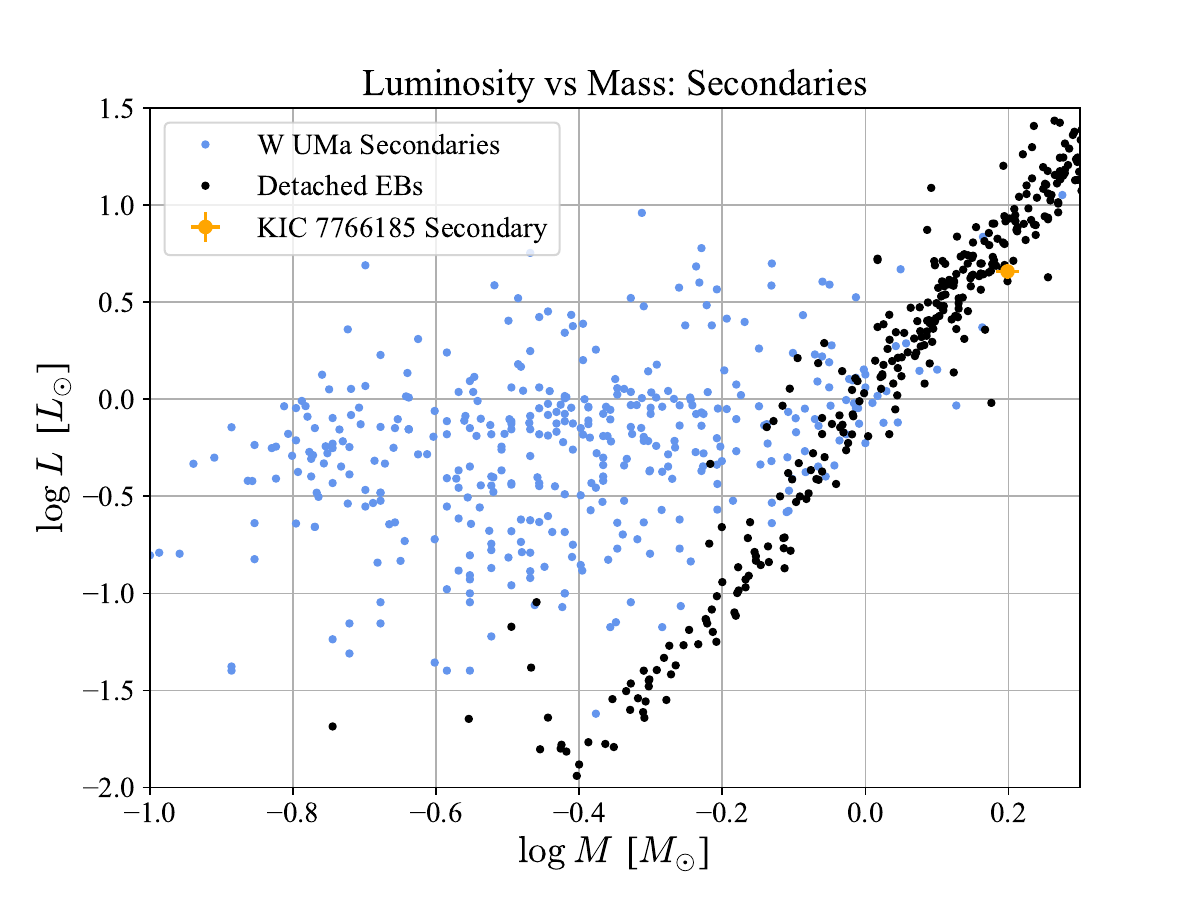}{0.4\textwidth}{(b)}}
    \caption{(a) Luminosity vs. Mass for detached EBs, W UMa primaries and the primary of KIC 7766185. (b) Luminosity vs. Mass for detached EBs, W UMa secondaries and the secondary of KIC 7766185.}
    \label{fig: HR}
\end{figure*}

\revIII{In figure \ref{fig: HR}a/b, we plot the luminosity vs. mass for the primaries/secondaries for W UMa stars compared to detached eclipsing binaries (EBs). The detached EB data is from \cite{HR_diagram}. Typically W UMa primaries are consistent with the detached EBs \citep{HR_explanation}. In the $\log L - \log M$ plane, the primary of KIC 7766185 is consistent with other W UMa primaries. The secondaries in W UMa systems are often displaced from the ZAMS and are inconsistent with the detached EBs. The secondaries are often over-luminous, which has been explained by energy transfer from the primary to the secondary \citep{HR_energy_transfer}. The secondary in KIC 7766185 does not appear over-luminous. It appears more consistent with the detached EBs rather than the other W UMa secondaries.}

\section{Discussion} \label{sec:discussion}\label{sec: discussion}

\begin{deluxetable*}{|l c c c|}
    \tablehead{\colhead{Parameter} & \colhead{This Study} & \colhead{\cite{Prsa_data}} & \colhead{\cite{other_study}}}
    \tablecaption{Comparison of Results}
    
     \startdata 
        $f$ & $0.029 \pm 0.002$ & 0.18943 & $0.08\pm0.03$ \\
        $i$ [$^{\circ}$] & $73.13^{+0.05}_{-0.04}$ & 77.79 & $72.4^{+1.1}_{-0.7}$ \\
        $q$ & $0.81 \pm 0.02$ & 0.81789 & $0.80^{+0.05}_{-0.04}$ \\
        $T_{2}/T_{1}$ & $0.9571^{+0.0006}_{-0.0007}$ & 0.97814 & $0.969^{+0.018}_{-0.019}$ \\
        $M_{1}$ [$M_{\odot}$]& $1.97\pm0.06$ & - & $1.76^{+0.13}_{-0.14}$ \\
     \enddata
    \tablecomments{Comparison of parameters determined in this study with those determined in \cite{Prsa_data} and \cite{other_study}.}
    \label{Tab: parameter commparison}
\end{deluxetable*}

This study conducted an in-depth analysis of KIC 7766185 using both photometric and spectroscopic observations. This object has been examined by previous works. Table \ref{Tab: parameter commparison} compares parameters determined in this study with those determined in \cite{Prsa_data} and \cite{other_study}. There are significant differences between the results of this study and \cite{Prsa_data}, particularly for the fillout factor and inclination. However, \cite{Prsa_data} utilized the EBAI (Eclipsing Binaries via Artificial Intelligence) estimator \citep{EBAI}, which utilizes artificial intelligence to estimate system parameters with light curves alone. Thus, it is unsurprising that this study, which solved both light curves and radial velocity curves with MCMC sampling yielded different results. Our determinations are consistent within error ranges to the \cite{other_study} results with the exception of the fillout factor and primary mass (and by extension, the secondary mass). \cite{other_study} rely on only two spectroscopic measurements at phases of 0.196 and 0.822. Additionally, they place a uniform prior on the fillout factor extending from 0.03 - 0.99. We find the fillout factor to be $0.029\pm0.002$ therefore the inclusion of this prior could be a cause of the discrepancies. Through examining the ETV curves of a sample of 253 short period binaries observed by both \textit{Kepler} and \textit{TESS}, \cite{other_period_study} have claimed a third body solution for KIC 7766185. Their proposed period (1886 days) is much shorter than the period used for the third body solution in Figure \ref{fig: ETV curve}. As mentioned earlier, while a third body solution is possible, further observations are needed to accurately determine any properties of the cyclic ETV variation (if the cyclic behavior is confirmed).

From the binary analysis, the sub-type of KIC 7766185 is determined. With a temperature difference of $T_{1} - T_{2} \approx 235$ K, we can rule out KIC 7766185 as a B-type system. The two stars are in geometric and thermal contact. With a temperature ratio of $T_{2}/T_{1} = 0.96$ and a radius radio of $R_{2}/R_{1} = 0.91$, the primary is the larger and hotter component, classifying KIC 7766185 as an A-type system. Additionally, with a mass ratio of $q = 0.81$, KIC 7766185 is an H-type system, which is the sub-set of high mass ratio W UMa stars with $q > 0.72$.

The results of this study and the comparison to other well studied W UMa systems allows us to speculate on the evolutionary state of KIC 7766185. This is a long period, high mass ratio and low fillout factor system. These last two characteristics are rare for A-type systems, which typically have smaller mass ratios and larger fillout factors \citep{Csizmadia, 700_WUMas}. \revIII{This may be a system which has only recently begun interacting. In the past, KIC 7766185 would have been a detached system. Through orbital decay (possibly driven by tidal dissipation and/or magnetic braking), the stars drifted closer towards one another until entering a semi-detached state where mass transfers from the primary to the secondary, which drives the mass ratio towards unity. For W UMa systems with some angular momentum loss mechanism, once contact is achieved, the system evolves to low mass ratios \citep{W_UMa_low_q}. If contact was achieved only recently, it could explain the low fillout factor and high mass ratio. The fact that the secondary is not over-luminous could be a consequence of the large secondary mass. The strong over-luminous secondary feature is prominent for low mass secondaries. The high mass secondaries are more consistent with the detached eclipsing binaries (see figure \ref{fig: HR}). It has been suggested that for the high mass ratio systems (the H-type systems), energy transfer from the primary to the secondary is less efficient \citep{Csizmadia}. If the components were born with similar masses, then the similar properties of the stars prior to interaction coupled with inefficient energy transfer could lead to neither component appearing over-luminous. This is just speculation however, detailed evolution models are required to better understand the evolutionary history and explain the unusual characteristics of KIC 7766185.}

The cyclic ETV variation (if confirmed) could indicate the presence of a third body or magnetic cycles \citep{Borkovits}. Either could drive the angular momentum loss of the system; a third body could drive tidal dissipation through Kozai-Lidov cycles or magnetic cycles could indicate the presence of a magnetic field which drives a magnetized stellar wind, causing angular momentum loss through magnetic braking.

\vspace{5 mm}

\section{Conclusion} \label{sec:conc}

We conducted a photometric and spectroscopic study of KIC 7766185. Spectra obtained from the Mayall 4-m telescope were used to extract radial velocity measurements. The radial velocity curve and the \textit{Kepler} light curve were used in \texttt{PHOEBE} to perform binary analysis which determined the stellar and orbital parameters. \textit{Gaia} multi-color photometry was also used to determine the effective temperatures of the two components. The system parameters confirm the system as an A-type W UMa star and reveal one of the largest and most massive secondary stars. The high mass ratio and very low fillout factor are rare for A-type systems. The period analysis determined a potential cyclical ETV variation which could be interpreted as a third body or magnetic cycles. However, future observations are required to confirm this cyclic behavior and accurately determine third body properties. Using the results of this study and comparisons to other well studied W UMa systems, we speculate that KIC 7766185 has only recently come in contact, having spent most of its evolution as a detached binary undergoing orbital decay through angular momentum loss.

C.L. graciously acknowledges support from the Delaware Space Grant College and Fellowship Program (NASA Grant 80NSSC20M0045). A.P. graciously acknowledges support from the National Science Foundation (AST \#2306996). The authors would also like to thank Dr. Jerome Orosz for the spectroscopic reduction process and Dr. Matthias Fabry for their informative insight.

\bibliography{Bibliography}{}
\bibliographystyle{aasjournal}

\end{document}